\documentclass[aps,pre,superscriptaddress,reprint, amsmath,amssymb,floatfix]{revtex4-1}
\usepackage{graphicx}
\usepackage{bm}
\usepackage{siunitx}
\usepackage{balance}
\begin{document}
	\title{Stress anisotropy in shear-jammed packings of frictionless disks}

	\author{Sheng Chen}
	%\email[]{wyan@flatironinstitute.org}
	\affiliation{Key Laboratory for Thermal Science and Power
  Engineering of Ministry of Education, Department of Energy and Power
  Engineering, Tsinghua University, Beijing 100084, China.}
  \affiliation{Department of Mechanical Engineering \& Materials Science,
Yale University, New Haven, Connecticut 06520, USA}
\author{Weiwei Jin}
%\email[]{wyan@flatironinstitute.org}
\affiliation{Department of Mechanics and Engineering Science, Peking     University, Beijing 100871, China.}
\affiliation{Department of Mechanical Engineering \& Materials Science,
Yale University, New Haven, Connecticut 06520, USA}

\author{Thibault Bertrand}
%\email[]{wyan@flatironinstitute.org}
\affiliation{Laboratoire Jean Perrin UMR 8237 CNRS/UPMC, Universit\'e Pierre et Marie Curie, Paris Cedex, 75255, France.}

\author{Mark D. Shattuck}
\affiliation{Benjamin Levich Institute and Physics Department,
The City College of the City University of New York, New York, New York 10031, USA.}

\author{Corey S. O'Hern}
\email[]{corey.ohern@yale.edu}
\affiliation{Department of Mechanical Engineering \& Materials Science,
Yale University, New Haven, Connecticut 06520, USA. E-mail: corey.ohern@yale.edu}
\affiliation{Department of Physics, Yale University, New Haven, Connecticut 06520, USA.}
\affiliation{Department of Applied Physics, Yale University, New Haven, Connecticut 06520, USA.}
	\date{\today}

	\begin{abstract}
    We perform
    computational studies of repulsive, frictionless disks to
    investigate the development of stress anisotropy in mechanically
    stable (MS) packings. We focus on two protocols for generating MS
    packings: 1) isotropic compression and 2) applied simple or pure
    shear strain $\gamma$ at fixed packing fraction $\phi$. MS packings
    of frictionless disks occur as geometric families (i.e. parabolic
    segments with positive curvature) in the $\phi$-$\gamma$ plane. MS
    packings from protocol 1 populate parabolic segments with both signs
    of the slope, $d\phi/d\gamma >0$ and $d\phi/d\gamma <0$.  In
    contrast, MS packings from protocol 2 populate segments with
    $d\phi/d\gamma <0$ only. For both simple and pure shear, we derive a
    relationship between the stress anisotropy and dilatancy
    $d\phi/d\gamma$ obeyed by MS packings along geometrical families.
    We show that for MS packings prepared using isotropic compression,
    the stress anisotropy distribution is Gaussian centered at zero with
    a standard deviation that decreases with increasing system size. For
    shear jammed MS packings, the stress anisotropy distribution is a
    convolution of Weibull distributions that depend on strain, which
    has a nonzero average and standard deviation in the large-system
    limit. We also develop a framework to calculate the stress
    anisotropy distribution for packings generated via protocol 2 in
    terms of the stress anisotropy distribution for packings generated
    via protocol 1.  These results emphasize that for repulsive
    frictionless disks, different packing-generation protocols give rise
    to different MS packing probabilities, which lead to differences in
    macroscopic properties of MS packings.
		%The abstrast goes here instead of the text "The abstract should be..."
	\end{abstract}
	% The abstract should be a single paragraph which summarises the content of the article. Any references in the abstract should be written out in full \textit{e.g.} [Surname \textit{et al., Journal Title}, 2000, \textbf{35}, 3523].

%Please use \dag to cite the ESI in the main text of the article.
%If you article does not have ESI please remove the the \dag symbol from the title and the footnotetext below.
%\footnotetext{\dag~Electronic Supplementary Information (ESI) available: [details of any supplementary information available should be included here]. See DOI: 10.1039/b000000x/}
%additional addresses can be cited as above using the lower-case letters, c, d, e... If all authors are from the same address, no letter is required

%\footnotetext{\ddag~Additional footnotes to the title and authors can be included \textit{e.g.}\ `Present address:' or `These authors contributed equally to this work' as above using the symbols: \ddag, \textsection, and \P. Please place the appropriate symbol next to the author's name and include a \texttt{\textbackslash footnotetext} entry in the the correct place in the list.}

%%%END OF FOOTNOTES%%%
	\maketitle
%%%MAIN TEXT%%%%
\section{Introduction}
\label{intro}

For systems in thermal equilibrium, such as atomic and molecular
liquids, macroscopic quantities, such as the shear stress and pressure,
can be calculated by averaging over the microstates of the system
weighted by the probabilities for which they occur, as determined by
Boltzmann statistics~\cite{Mcquarrie2000}.  In contrast, granular
materials, foams, emulsions, and other athermal particulate media are
out of thermal equilibrium and this formalism breaks
down~\cite{behringer,makse}.

For dense, quasistatically driven particulate media, the relevant
microstates are mechanically stable (MS) packings with force- and
torque-balance on all grains~\cite{gao1,ShenPRL2014}. In contrast to
thermal systems, the probabilities with which MS packings occur are
highly non-uniform and depend on the protocol that was used to
generate them~\cite{gao2}. For example, it has been shown that MS
packings generated via vibration, compression, and pure and simple
shear possess different average structural and mechanical
properties~\cite{MajmudarNature2005,BiNature2011,BertrandPRE2016}. In
previous work on jammed packings of purely repulsive frictionless
disks, we showed that the differences in macroscopic properties do not
occur because the collections of microstates for each protocol are
fundamentally different, instead the probabilities with which
different MS packings occur change significantly with the
protocol~\cite{BertrandPRE2016}. Thus, it is of fundamental importance
to understand the relationship between the packing-generation protocol
and MS packing probabilities.

\begin{figure}[h]
  \includegraphics[width = 8.5 cm]{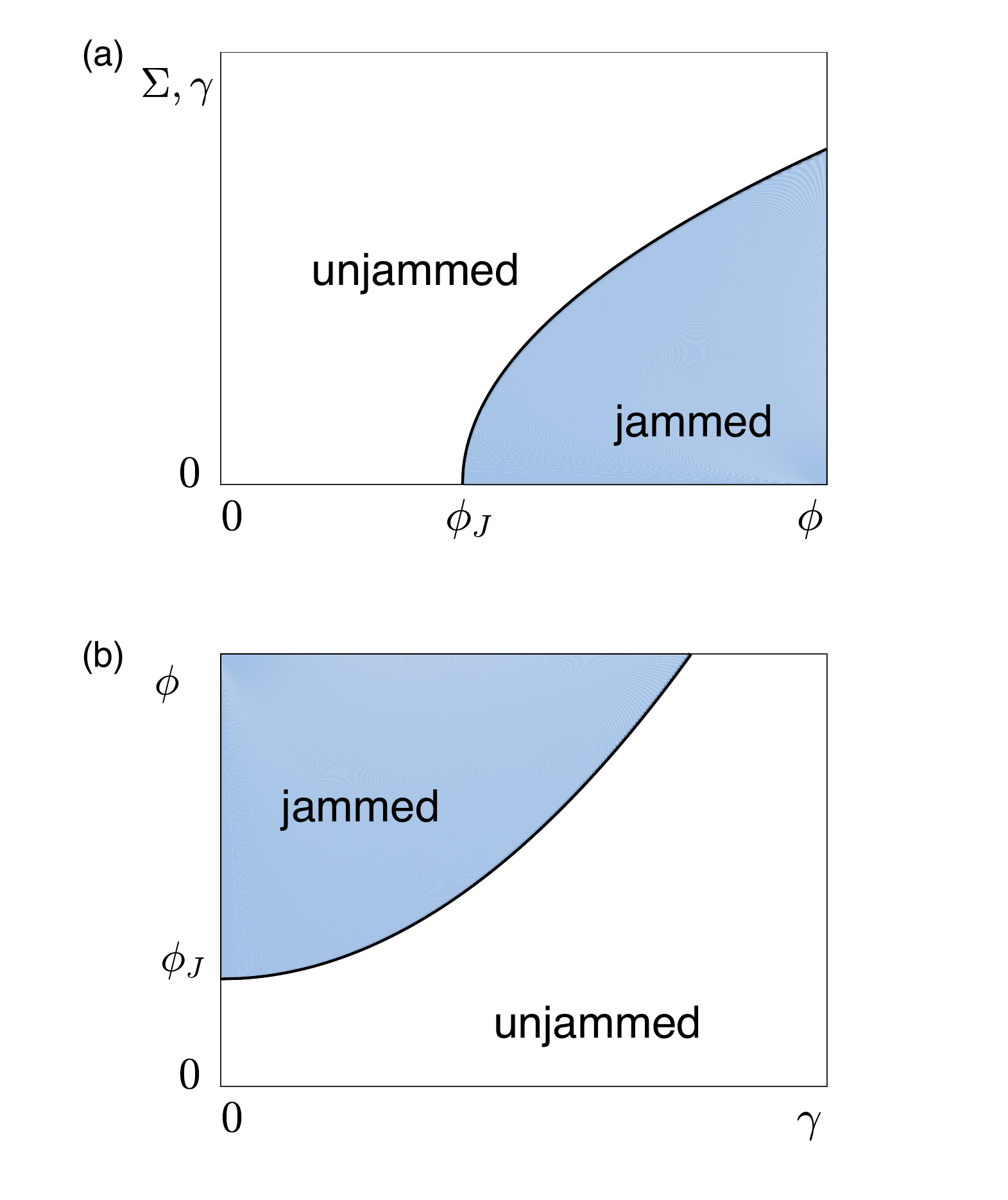}
\caption{(a) A schematic jamming phase diagram in the
stress $\Sigma$ and packing
fraction $\phi$ plane. The solid line indicates the yield stress
$\Sigma^y(\phi)$. For applied stress $\Sigma < \Sigma^y$, the system is
jammed and for $\Sigma > \Sigma^y$, the system flows and is unjammed.
We assume that the yield strain $\gamma^y$ scales with the yield stress and
obeys $\Sigma^y \sim
\gamma^y \sim (\phi - \phi_J)^{\nu}$, where $\nu = 0.5$ and $\phi_J$
is the jammed packing fraction in the absence of shear stress. (b) The
same
jamming phase diagram in (a) except rendered in the $\phi$-$\gamma$ plane. The jammed packing
fraction increases quadratically with strain
from $\phi_J$. With the phase diagrams in (a) and (b), increasing
strain does not cause a system to transition from unjammed to jammed.}
\label{fig1}
\end{figure}

Jamming, where an athermal particulate system transitions from a
liquid-like to a solid-like state with a non-zero yield stress,
induced by isotropic compression has been studied in granular and
other athermal materials for more than 20
years~\cite{MajmudarNature2005,clusel,ohern}. Recently, Bi, {\it et
  al.} showed that packings of granular disks can jam via simple
and pure shear at fixed area~\cite{BiNature2011}.  This was a
surprising result because many previous studies had emphasized that
the application of shear at fixed packing fraction gives rise only to
flow and unjamming behavior.  This point is emphasized in the
schematic jamming phase diagram in the stress $\Sigma$ and packing
fraction $\phi$ plane in Fig.~\ref{fig1} (a), which shows that the
yield stress $\Sigma^y$ or strain $\gamma^y$ increases with $\phi$
above jamming onset $\phi_J$ at zero shear. Here, we assume that
$\Sigma^y \sim \gamma^y \sim (\phi - \phi_J)^{\nu}$, where $\nu =
0.5$. In Fig.~\ref{fig1} (b), we flip the axes so that the packing
fraction at the yield strain increases quadratically from $\phi_J$
with increasing strain. In this picture, increasing
the shear strain does not give rise to jamming.  However, we will show
below that this picture is incomplete, and the application of shear
strain can cause unjammed systems of frictionless, spherical particles
to jam~\cite{BertrandPRE2016,luding}.

Despite important work~\cite{baity,BertrandPRE2016,luding} since the
original manuscript by Bi, {\it et al.}, there are still many open
questions concerning shear jamming. For example, 1) Can shear jamming
occur in MS packings of frictionless grains and if so, do these
shear-jammed packings possess a nonzero stress anisotropy? and 2) Are
there substantive differences between MS packings generated via
isotropic compression versus shear?

Our recent work has shown that mechanically stable packings of
frictionless spherical particles can be obtained via either simple
shear or isotropic compression and that the probability for a
particular packing depends on the packing-generation
protocol~\cite{BertrandPRE2016}. The average shear strain required to
jam an originally unjammed configuration can be written in terms of
the basin volume, density of jammed packings, and path in
configuration space from the initial condition to the final MS
packing. This previous work focused mainly on the shear strain
$\gamma_{J}$ needed to jam an initially unjammed configuration and how
the shear strain $\gamma_J$ depends on the packing fraction. In the
current article, we instead focus on the shear stress anisotropy in MS
packings generated by isotropic compression versus pure and simple
shear.

Our computational studies yield several key results, which form a more
complete picture of shear jamming in packings of frictionless
spherical particles. First, we identify relationships between the
stress anisotropy and the packing fraction and its derivative with
respect to strain (dilatancy) for MS packings generated via simple and
pure shear. These relationships allow us to calculate the stress
anisotropy (which includes contributions from both the shear stress
and normal stress difference) for MS packings by only knowing how the
jammed packing fraction varies with strain.  Second, we show that the
distribution of the stress anisotropy for isotropically compressed
packings is a Gaussian centered on zero with a width that decreases as
a power-law with increasing system size $N$~\cite{xu}. In contrast,
the stress anisotropy distribution is a convolution of
strain-dependent Weibull distributions with a finite average and
standard deviation in the large-system limit for shear-jammed MS
packings~\cite{abe}.  Fourth, using the relation between stress
anisotropy and dilatancy, we predict the stress anisotropy
distribution for shear-jammed packings using that for MS packings
generated via isotropic compression.

The remainder of the article includes three sections and three
appendices, which provide additional details to support the
conclusions in the main text. In Sec.~\ref{methods}, we describe the
two main protocols that we use to generate MS packings and provide
definitions of the stress tensor and stress
anisotropy. Sec.~\ref{results} includes four subsections that
introduce the concept of geometrical families, derive the
relationships between the stress tensor components and the dilatancy,
develop a framework for calculating the shear stress distribution for
shear-jammed packings in terms of the shear stress distribution for
isotropically compressed packings, and describe the robustness of our
results are with increasing system size. In Sec.~\ref{conclusions}, we
give our conclusions, as well as describe interesting future
computational studies on shear-jammed packings of non-spherical
particles, such as circulo-polygons~\cite{vanderwerf}, and frictional
particles~\cite{ShenPRL2014}.

\section{Methods}
\label{methods}

Our computational studies focus on systems in two spatial dimensions
containing $N$ frictionless bidisperse disks that interact via the
purely repulsive linear spring potential given by $V(r_{ij}) =
  \frac{\epsilon}{2} (1-r_{ij}/\sigma_{ij})^2
  \Theta(1-r_{ij}/\sigma_{ij})$, where $\epsilon$ is the strength of
the repulsive interactions, $r_{ij}$ is the separation between the
centers of disks $i$ and $j$, $\sigma_{ij} = (\sigma_i+\sigma_j)/2$,
$\sigma_i$ is the diameter of disk $i$, and $\Theta(.)$ is the
Heaviside step function that prevents non-overlapping particles from
interacting.  The system includes half large disks and half small
disks with diameter ratio $r=1.4$. The disks are confined within an
undeformed square simulation cell with side lengths, $L_x = L_y = 1$,
in the $x$- and $y$-directions, respectively, and periodic boundary
conditions. Isotropic compression is implemented by changing the cell
lengths according to $L_x' = L_x(1-d \phi/2\phi)$ and $L_y'
  =Ly(1-d \phi/2\phi)$ and corresponding affine shifts in the
particle positions, where $d \phi < 10^{-4}$ is the change in packing
fraction. Simple shear strain with amplitude $\gamma$ is implemented
using Lees-Edwards periodic boundary conditions, where the top
(bottom) images of the central cell are shifted to the right (left) by
$\gamma L_{y}$ with corresponding affine shifts of the particle
positions~\cite{LeesJOP1972}. Pure shear is implemented by compressing
the simulation cell along the $y$-direction and expanding it along the
$x$-direction with corresponding affine shifts of the particle
positions. The system area is kept constant (i.e. $A=L_x'L_y' = L_x
L_y$) and the pure shear strain is defined as $\gamma =
\ln(L_x'/L_y')$.

%\subsection{2.2 Two protocols to generate jammed packings}

\begin{figure}
\includegraphics[width=8.3 cm]{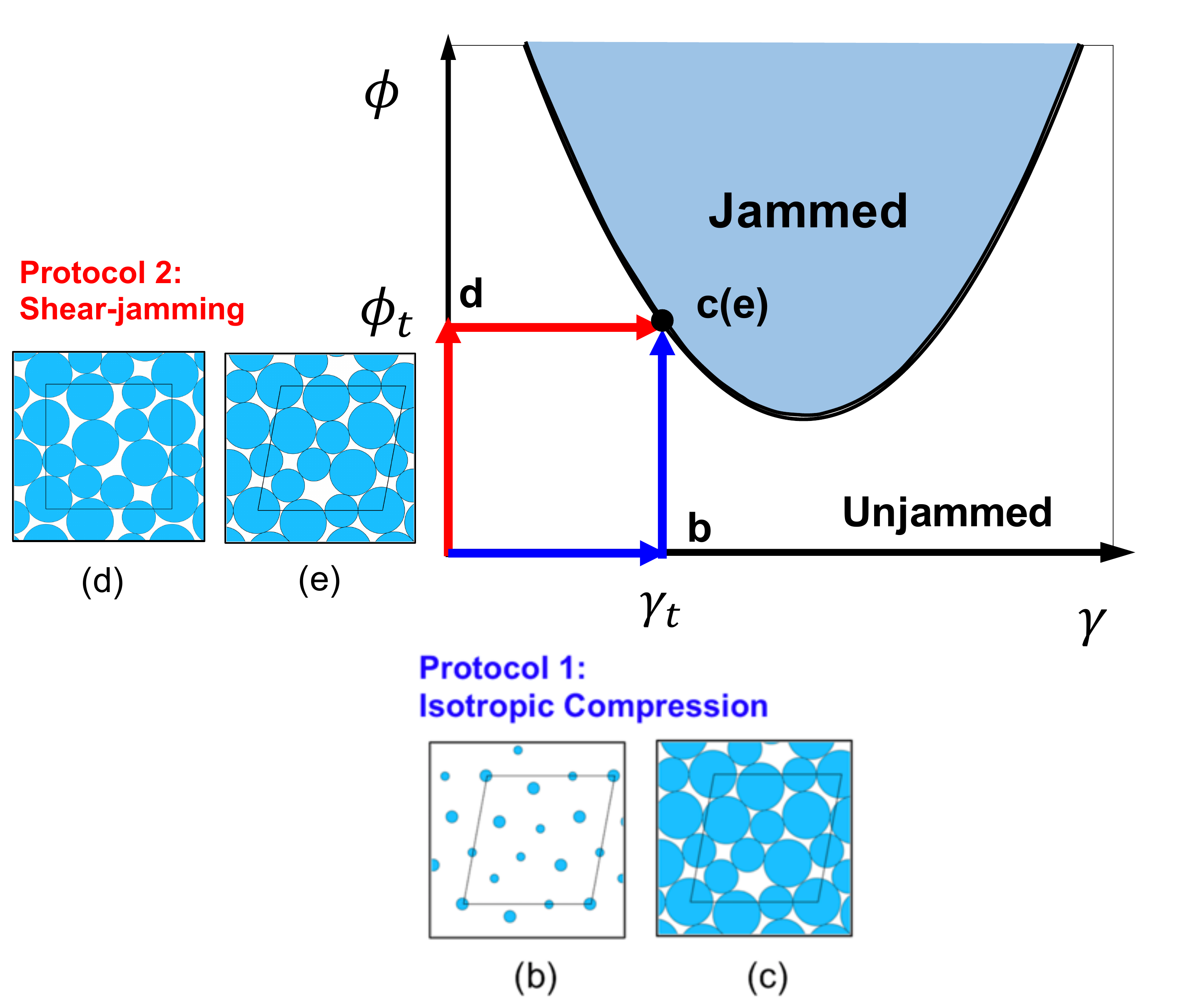}
\caption{\label{fig1_protocol} Schematic of the packing fraction
$\phi$ and simple shear strain $\gamma$ plane that illustrates the two main
protocols used to generate MS disk packings. As shown in
Fig.~\ref{fig2_fam} (e), the jammed regions are bounded by parabolic
segments.
In protocol 1, the system
is first deformed to simple shear strain $\gamma_t$ at small
initial packing fraction
$\phi_i \approx  0$ (point b) and then isotropically compressed to
jamming onset at $\phi_t$ (point c). In protocol 2, the system is first
compressed to
$\phi_t$ below jamming onset (point d) at $\gamma_i = 0$ and then sheared to
jamming onset at simple shear strain $\gamma_t$
(point e). Points (c) and (e) correspond to the same total deformation,
and thus the two
protocols can yield the same MS packing. Note that for each system size $N$, there are many distinct parabolas that
occur over a range of strain and packing fraction. As the system size
increases, the typical parabolic segment size decreases as $1/N$ and
the range of packing fraction over which the parabolic segments occur
shrinks to zero.}
\end{figure}

As shown in Fig.~\ref{fig1_protocol}, we employ two main protocols to
generate MS packings in the packing fraction $\phi$ and shear strain
$\gamma$ plane. For protocol $1$, we first place the disks at random
initial positions in the simulation cell, and apply successive simple
shear strain steps $d \gamma < 10^{-4}$ to total strain $\gamma_t$ at
fixed small packing fraction $\phi_i = 0.1$. We then
isotropically compress the system in small packing fraction increments
$d\phi$ to jamming onset $\phi_J$ at fixed simple shear strain
$\gamma=\gamma_t$. For protocol 2, we first place the disks at random
initial positions and then isotropically compress the system to a
target packing fraction $\phi_{t} < \phi_J$ at simple shear strain
$\gamma_i=0$. We then apply simple shear to the system in small
strain steps $d \gamma$ until the system jams at $\gamma_J$. For
protocol 2, the target volume fraction $\phi_t$ varies from $\phi_m$,
below which no shear-jammed packings can be found in the range $0 <
\gamma < 1$ to $\phi_J$ obtained from isotropic compression at
$\gamma=0$.  In Appendix A, we also include results for a
packing-generation protocol similar to protocol 2, except we apply
pure instead of simple shear strain.

The total potential energy per particle $U=U'/N\epsilon$, where
$U'=\sum_{i>j} V(r_{ij})$, is minimized using the conjugate gradient
technique after each compression or shear step. Minimization is
terminated when the potential energy difference between successive
conjugate gradient steps satisfies $\Delta U/U < 10^{-16}$. We define
jamming onset when the total potential energy per particle obeys
$U_{\rm max}<U<2U_{\rm max}$, with $U_{\rm max} =
10^{-16}$. This method for identifying jamming onset is similar to that
used in our previous studies~\cite{BertrandPRE2016}.

The systems are decompressed (for protocol 1) or sheared in the
negative strain direction (for protocol 2) when $U$ at a local minimum
is nonzero, i.e., there are finite particle overlaps. If the potential
energy is “zero” (i.e. $U < 10^{-16}$), the system is compressed (for
protocol 1) or sheared in the positive strain direction (for protocol
2). For protocol 1, the increment by which the packing fraction is
changed at each compression or decompression step is halved each time
$U$ switches from zero to nonzero or vice versa. Similarly, for
protocol 2, the increment by which the shear strain is changed at each
strain step is halved each time $U$ switches from zero to nonzero or
vice versa. These packing-generation protocols yield mechanically
stable packings (with a full-spectrum of nonzero frequencies of the
dynamical matrix~\cite{barrat}) at jamming onset. In addition, all of
the MS disk packings generated via protocols 1 and 2 are isostatic,
where the number of contacts matches the number of degrees of freedom,
$N_c = N_c^0$, with $N_c^0 = 2N'-1$, $N' = N-N_r$, and $N_r$ is the
number of rattler disks with fewer than three contacts~\cite{witten}.

%fig 1

For each MS packing, we calculate the stress tensor:
%equation (1)
\begin{equation}
\label{eq_stress}
\Sigma_{\beta \delta}=\frac{1}{A}\sum_{i \neq j}f_{ij\beta}r_{ij\delta},
\end{equation}
where $A=L_x L_y$ is the system area, $f_{ij\beta}$ is the
$\beta$-component of the interparticle force on particle $i$ due to
particle $j$, $r_{ij\delta}$ is the $\delta$-component of the
separation vector from the center of particle $j$ to that of particle
$i$, and $\beta$ and $\delta=x$,$y$. From the components of the stress
tensor, we can calculate the pressure $P=(\Sigma_{xx}+\Sigma_{yy})/2$,
the normal stress difference $\Sigma_N = (\Sigma_{yy}-\Sigma_{xx})/2$,
and the shear stress $-\Sigma_{xy}$.  We define the normalized stress
anisotropy to be $\hat{\tau} = \sqrt{\hat{\Sigma}_N^2 +
  \hat{\Sigma}_{xy}^2}$, where $\hat{\Sigma}_{N}=\Sigma_{N}/P$ and
$\hat{\Sigma}_{xy} = -\Sigma_{xy}/P$. $\hat{\tau}$ includes
contributions from both the shear stress and the normal stress
difference. We will show below that only the shear stress (normal
stress difference) contributes to $\hat{\tau}$ for MS packings
generated via simple shear (pure shear). Therefore, we will focus on
$\hat{\Sigma}_{xy}$ when we study packings generated via simple shear
and on $\hat{\Sigma}_{N}$ when we study packings generated via pure
shear.  (See Appendix A.)  We calculate mean values and standard
deviations of the stress tensor components over between $10^3$ and
$10^5$ distinct MS packings.

\section{Results}
\label{results}

\subsection{Geometrical families}
\label{families}

\begin{figure}[h]
\includegraphics[width=8.8 cm]{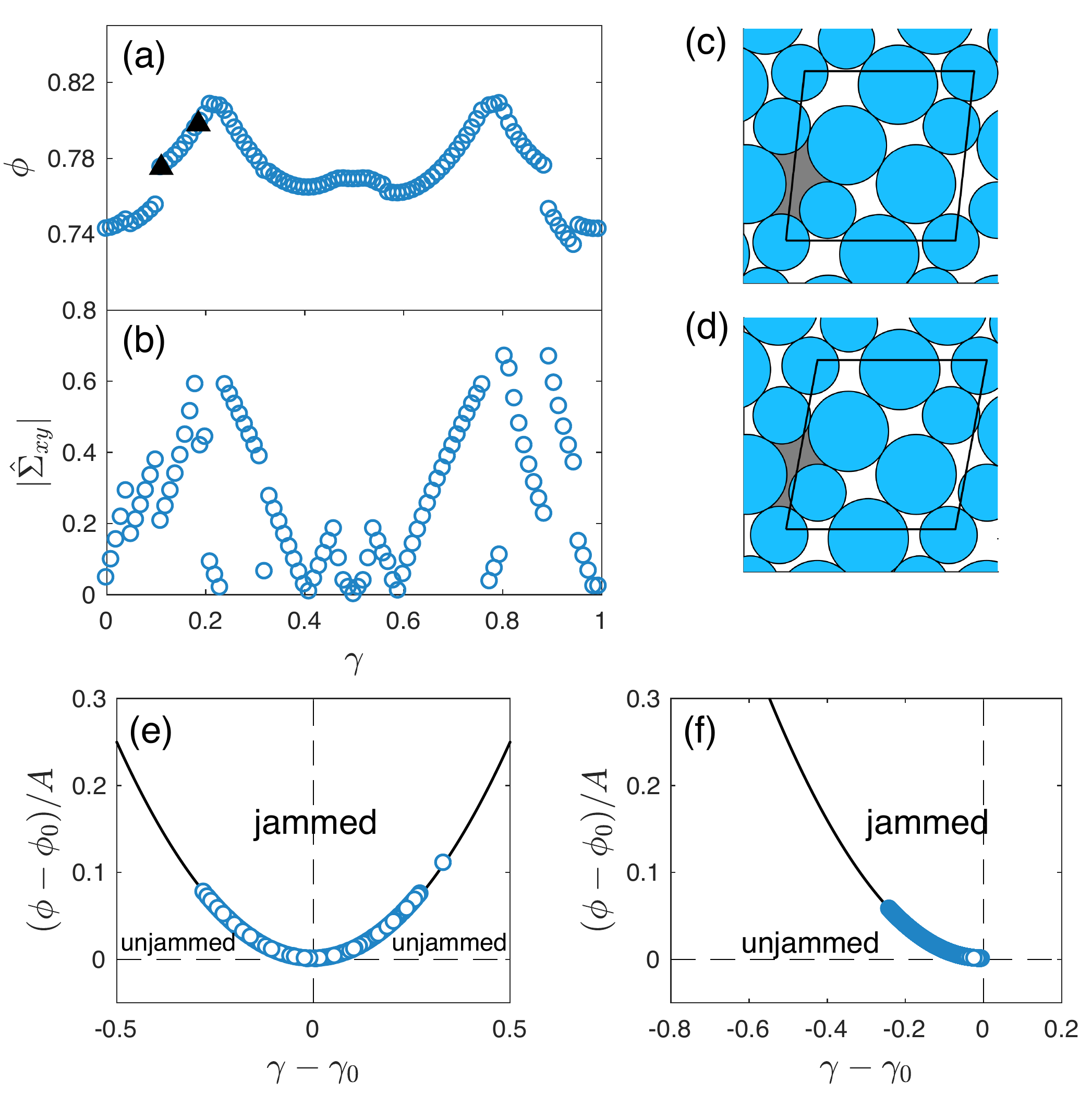}
\caption{\label{fig2_fam} (a) Packing fraction $\phi$ at jamming onset
as a function of simple shear strain $\gamma$ for MS packings with $N=6$
generated via isotropic compression (protocol 1) and (b) the corresponding
magnitude of the shear
stress $\left|\hat{\Sigma}_{xy}\right|$ versus $\gamma$. The data in
(a) and (b) were
obtained using the same single set of random initial conditions. Panels (c) and (d) show the
MS packings near the start and the end of a geometrical family, indicated
by the lower and upper filled triangles in (a). Each geometrical family in (a), as well as
the families obtained from other random initial conditions, can be described
by parabolic segments, $\phi = A(\gamma-\gamma_0)^2 + \phi_0$, in the $\phi$-$\gamma$
plane, where
$A>0$, $\phi_0$, and $\gamma_0$ are the curvature, packing fraction offset,
and strain offset for each geometrical family.
Panels (e) and (f) show the normalized coordinates, $(\phi-\phi_0)/A$ versus
$\gamma-\gamma_0$, for all MS packings with $N=6$ generated via protocols 1 and 2,
respectively. Protocol 1 generates packings with both signs of
$d\phi/d\gamma$, whereas protocol 2 only generates packings
with $d\phi/d\gamma < 0$. The jammed and unjammed regions of the $(\phi-\phi_0)/A$ and $\gamma-\gamma_0$ plane are indicated.}
\end{figure}

As background, we review the structure of geometrical families during
shear deformation~\cite{GaoPRE2009,BertrandPRE2016}.  In
Fig.~\ref{fig2_fam} (a), we illustrate that MS packings occur as
geometrical families, forming continuous segments in the jammed
packing fraction $\phi$ and shear strain $\gamma$ plane, with the same
interparticle contact networks. In panel (a), the $N=6$ MS packings
were generated using isotropic compression (protocol 1) from a single
random initial condition. In Fig.~\ref{fig2_fam} (c) and (d), we
highlight two MS packings near the beginning and end of the
geometrical family indicated by the filled triangles in (a). The
system switches from one geometrical family to another when the
interparticle contact network becomes unstable. The beginning and end
of each geometrical family can be identified by finding changes in the
interparticle contact network or discontinuous changes in
$\phi(\gamma)$ or slope $d\phi/d\gamma$.

Each geometrical family of MS packings forms a parabolic segment in the
$\phi$-$\gamma$ plane described by $\phi(\gamma) =
A(\gamma-\gamma_0)^2 + \phi_0$, where $A$, $\gamma_0$, and $\phi_0$
give the curvature, strain offset, and packing fraction offset for
each family. The curvature satisfies $A>0$ for all geometrical
families of MS disk packings. In Fig.~\ref{fig2_fam} (e) and (f), we
show that the data collapse onto a parabolic form when we plot
$(\phi-\phi_0)/A$ versus $\gamma-\gamma_0$ for all geometric families
we found using protocols 1 and 2, respectively, with more than $10^5$
initial conditions. For protocol 1, we obtain families with both
$d\phi/d\gamma >0$ and $d\phi/d\gamma < 0$. However, for protocol 2,
the geometrical families only possess $d\phi/d\gamma < 0$.  For
protocol 1, the systems approach the jammed region from below, and
thus they can reach both sides of the parabolas. For protocol 2, the
systems approach the jammed region from the left, and thus they jam
when they reach the left sides of the parabolas. Note the key
difference in the signs of the slope, $d\phi/d\gamma$, between the
jamming phase diagrams in Figs.~\ref{fig1} (b) and~\ref{fig2_fam} (f).
The schematic jamming phase diagram in Fig.~\ref{fig1} (b) is missing
the portion of the parabola with $d\phi/d\gamma < 0$.

The geometrical family structure can also be seen in the shear stress
versus strain as shown in Fig.~\ref{fig2_fam} (b).  In this case, the
shear stress $|{\hat \Sigma}_{xy}|$ varies quasi-linearly with
$\gamma$.  For MS packings within a given geometrical family, we find
that $|\hat{\Sigma}_{xy}|$ increases with $\phi$ and
$|\hat{\Sigma}_{xy}| \approx 0$ when $\phi(\gamma)$ is near a local
minimum or maximum (i.e., $\frac{\partial\phi}{\partial\gamma} = 0$).
Although we illustrated these results for a small system, we showed in
previous studies~\cite{BertrandPRE2016} that the geometrical family
structure persists with increasing system size. In large-system limit, the
family structure occurs over a narrow range of $\phi$ near $\phi_J
\approx 0.84$, and the system only needs to be sheared by an infinitesimal
strain to switch from one family to another.

%fig 2

\subsection{Relationship between the stress tensor components and dilatancy}
\label{dilatancy}

In this section, we derive relationships between the components of the
stress tensor (i.e. the shear stress ${\hat \Sigma}_{xy}$ and normal
stress difference ${\hat \Sigma}_N$) and the packing fraction and
dilatancy~\cite{PeyneauPRE2008,kabla,kruyt}, $d\phi/d\gamma$, for MS
packings generated via protocols 1 and 2.  For MS packings belonging
to a given geometrical family, the total energy does not change
following a strain step $d\gamma$ and a decompression step
that changes the area by $dA$. Thus,
the total work is given by $-PdA -\Sigma_{xy} A d\gamma = 0$ for
simple shear and $-P dA- \Sigma_{yy}L_x' dL_y' -
\Sigma_{xx}L_y'dL_x' = 0$ for pure shear.  Using $dA/A=-d\phi/\phi$,
we find
\begin{equation}
\label{simple_shear}
{\hat \Sigma}= -\frac{1}{\phi}\frac{d\phi}{d\gamma},
\end{equation}
where ${\hat \Sigma}={\hat \Sigma}_{xy}$ for simple shear and ${\hat \Sigma}_{N}$ for pure shear deformations.
Thus, the shear stress $\hat{\Sigma}_{xy}$ (normal stress difference
$\hat{\Sigma}_{N}$) along a geometrical family is proportional to the
dilatancy, $d\phi/d\gamma$, during simple (pure) shear deformation.

\begin{figure}
\includegraphics[width=8.4 cm]{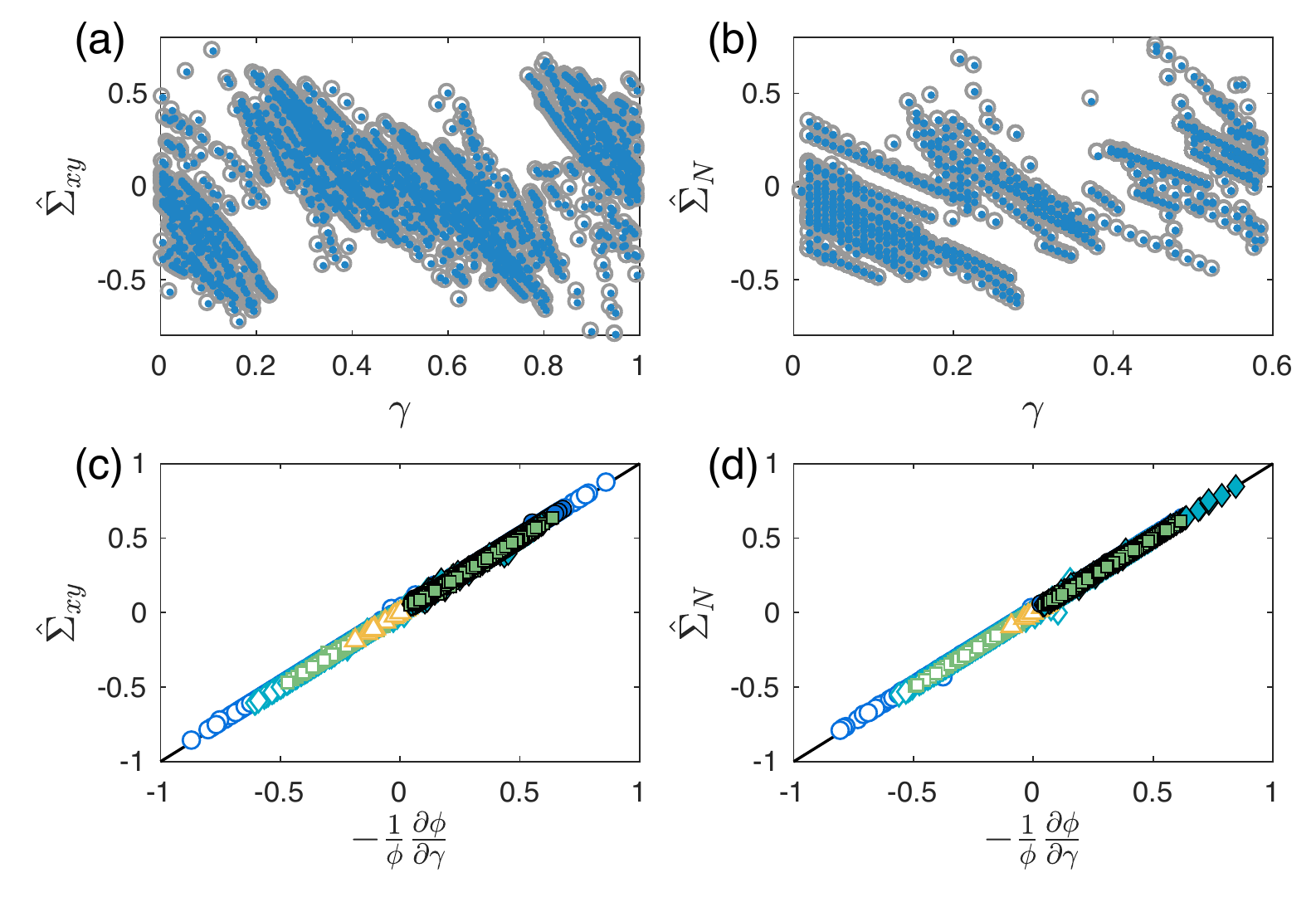}
\caption{\label{fig3_mapall} (a) Shear stress $\hat{\Sigma}_{xy}$ versus simple shear strain $\gamma$ and (b) normal stress difference $\hat{\Sigma}_{N}$ versus
pure shear strain $\gamma$ for $N=6$ MS packings generated via isotropic
compression (protocol 1). Gray circles are data points obtained from
the components of the stress tensor and
blue dots are obtained by finding all of the geometrical families
and calculating $\hat{\Sigma}_{xy}$ and $\hat{\Sigma}_{N}$ from
Eq.~\ref{simple_shear} along each family. Panels
(c) and (d) show plots of $\hat{\Sigma}_{xy}$ and $\hat{\Sigma}_{N}$
calculated using the stress tensor versus the results from
Eq.~\ref{simple_shear} for MS packings
with $N = 6$ (circles),
$10$ (diamond), $16$ (squares), and $32$ (upward triangles). Open
(solid) symbols indicate MS packings generated via protocol 1
(protocol 2). The solid line has unit slope and zero vertical intercept.}
\end{figure}

In Fig.~\ref{fig3_mapall} (a) and (b), we compare the results from the
calculations of the shear stress and normal stress difference using
the stress tensor (Eq.~\ref{eq_stress}) to those using
Eq.~\ref{simple_shear} %and~\ref{pure_shear}
for $N=6$ MS packings
generated using protocol 1.  We find strong agreement.  In
Fig.~\ref{fig3_mapall} (c) and (d), we further compare the two methods
for calculating the stress tensor components by plotting ${\hat \Sigma}_{xy}$ or ${\hat \Sigma_N}$ from the stress tensor versus the right side of Eq.~\ref{simple_shear} for several system sizes and protocols 1 and 2. The data collapse onto a line with unit slope and zero vertical intercept. Data points that deviate from the straight line collapse onto the line when $d\gamma$ is decreased to $2\times 10^{-4}$.

%fig3

\subsection{Distributions of the shear stress and normal stress difference
for protocols 1 and 2}
\label{stress_anisotropy}

\begin{figure}[h]
\includegraphics[width=8.0 cm]{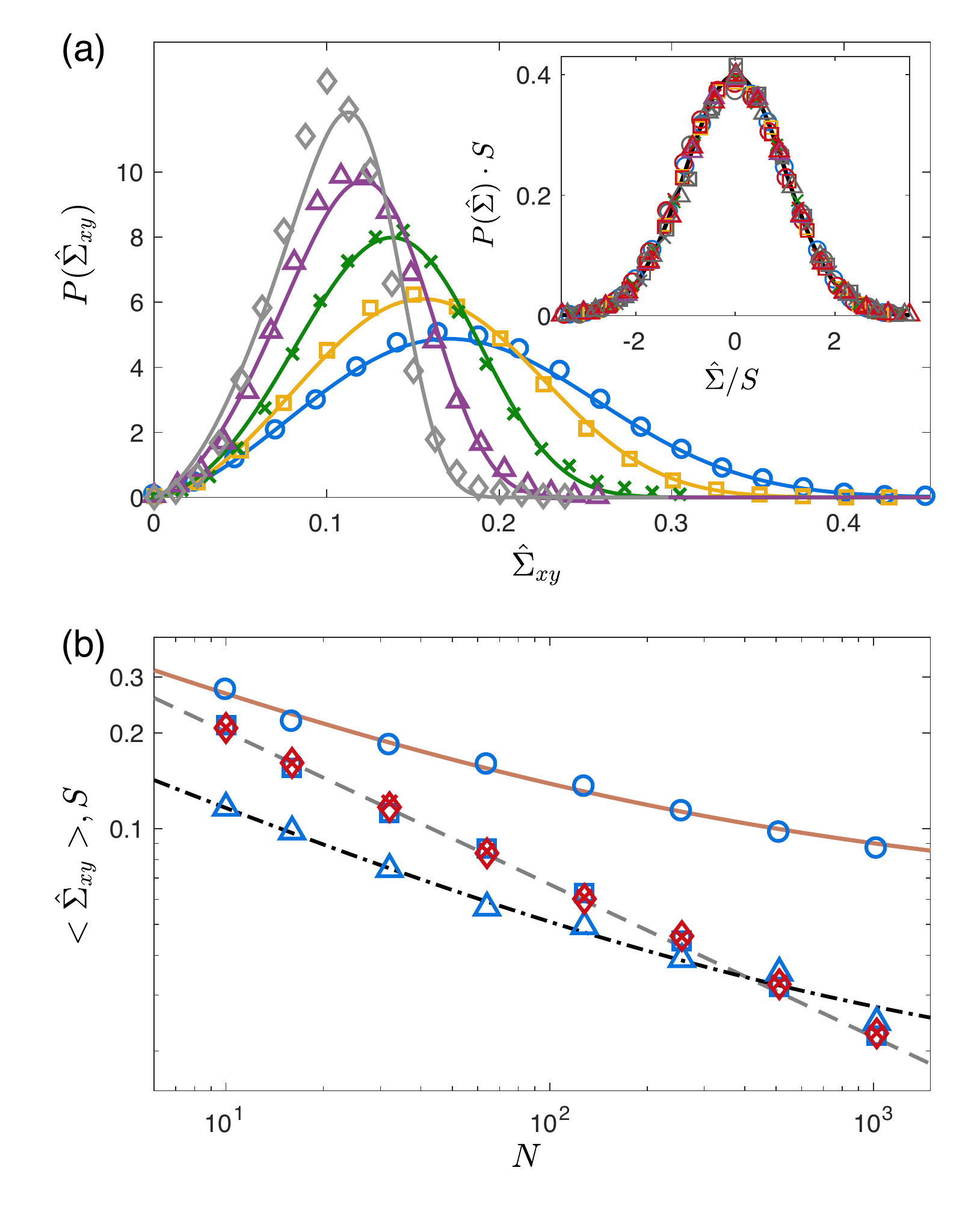}
\caption{\label{fig4_pdf}
(a) The probability distributions of the shear stress $P(\hat{\Sigma}_{xy})$
for MS packings generated via protocol 2 with simple shear for
$N = 32$ (circles), $64$ (squares), $128$ (crosses), $256$
(triangles), and $512$ (diamonds). The solid lines are predictions from
Eq.~\ref{eq_pdf_prediciton}. In the inset, we show three types of probability
distributions scaled by their standard deviations $S$:
$P(\hat{\Sigma}_{xy})$ (same symbols as main panel) and $P({\hat \Sigma}_N)$
(same symbols as main panel, but in red) for MS packings
generated via isotropic compression (protocol 1) and $P({\hat \Sigma}_N)$
for protocol 2 with simple shear (same symbols as main panel, but in gray).
The solid black line is a Gaussian distribution
with zero mean and unit standard deviation. (b)
System-size dependence of $\langle \hat{\Sigma}_{xy} \rangle$ (circles) and
standard deviations of $P(\hat{\Sigma}_{xy})$
(triangles) and $P(\hat{\Sigma}_{N})$ (squares) for MS
packings generated via protocol 2 with simple shear and the standard
deviations of $P(\hat{\Sigma}_{xy})$ (crosses)
and $P(\hat{\Sigma}_{N})$ (diamonds) for MS packings generated
via protocol 1. The dashed, solid, and dash-dotted lines are fits to
Eqs.~\ref{s_cale_p1},~\ref{eq_stress_rs2}, and~\ref{s_cale_p2}, respectively.}
\end{figure}

In the inset of Fig.~\ref{fig4_pdf} (a), we show the probability
distributions for the shear stress and normal stress difference,
$P(\hat{\Sigma}_{xy})$ and $P(\hat{\Sigma}_{N})$, for MS packings
generated via isotropic compression (protocol 1) and
$P(\hat{\Sigma}_{N})$ for MS packings generated via protocol 2 with
simple shear. When scaled by the standard deviation $S$, these
distributions collapse onto a Gaussian curve centered at zero with
unit standard deviation. As shown in Fig.~\ref{fig4_pdf} (b), the standard
deviations for all three distributions scale with system size as
%equation
\begin{equation}
\label{s_cale_p1}
S_1(N) = S^0_{1} N^{-\omega_1},
\end{equation}
where $S^0_1 \approx 0.61$ and $\omega_1 \approx 0.48$. Thus, the stress tensor
is isotropic in the large system-limit for MS packings generated via
isotropic compression (protocol 1).  In addition, the normal stress difference
is zero for MS packings generated via protocol 2 with simple shear.

%fig 4

\begin{table}[h]
  \centering
  \label{table}
  \caption{Means ($\langle .\rangle$) and standard deviations ($S$) of the shear stress ${\hat \Sigma}_{xy}$ and normal stress difference ${\hat \Sigma}_N$
distributions in the large-system limit for protocols 1 and 2.}
\begin{tabular}{l*{5}{c}r}
\hline
Protocol     & $\langle \hat{\Sigma}_{xy}\rangle_{\infty}$  &  $\langle \hat{\Sigma}_{N}\rangle_{\infty}$  &  $S^{xy}_{\infty}$        &  $S^{N}_{\infty}$    \\
\hline
Protocol 1                   &          0  &   0                & 0         &     0   \\
\hline
Protocol 2 \\
simple shear   & 0.060        &   0                & 0.015     &     0    \\
\hline
protocol 2  \\
 pure shear    &   0          &  0.055             & 0         &    0.016 \\
\hline
\end{tabular}
\end{table}

In the main panel of Fig.~\ref{fig4_pdf} (a), we show the probability
distribution of the shear stress $P({\hat \Sigma}_{xy})$ for MS
packings generated via protocol 2 with simple shear.  We note that
${\hat \Sigma}_{xy} > 0$ and $P({\hat \Sigma}_{xy})$ is non-Gaussian
for protocol 2. In contrast to the behavior of the average shear
stress $\langle {\hat \Sigma}_{xy} \rangle$ for MS packings generated
via isotropic compression (protocol 1), $\langle \hat{\Sigma}_{xy}
\rangle$ approaches a nonzero value in the large-system limit for MS
packings generated via protocol 2 with simple shear. As shown in
Fig.~\ref{fig4_pdf} (b),
\begin{equation}
\label{eq_stress_rs2}
\langle \hat{\Sigma}_{xy} \rangle(N) = \hat{\Sigma}_0 N^{-\Omega}+\hat{\Sigma}_{\infty},
\end{equation}
where $\hat{\Sigma}_0 \approx 0.54$, $\Omega \approx 0.42$, and
$\hat{\Sigma}_{\infty} \approx 0.060$. Similarly, we find that the
standard deviation of $P({\hat \Sigma}_{xy})$ for MS packings
generated via protocol 2 with simple shear approaches a nonzero value
in the large-system limit:
\begin{equation}
\label{s_cale_p2}
S_2(N) = S_2^0 N^{-\omega_2}+S_{\infty},
\end{equation}
where $S_2^0 \approx 0.28$, $\omega_2 \approx 0.45$, and $S_{\infty}
\approx 0.015$. In contrast, the width of the distribution of
jammed packing fractions tends to zero in the large-system limit~\cite{ohern}.
Thus, the packing-generation protocol strongly
influences the stress anisotropy, especially in the large-system
limit. The results for the average values and standard deviations of
the distributions $P({\hat \Sigma}_{xy})$ and $P({\hat \Sigma}_{N})$ in
the large-system limit for protocols 1 and 2 (for simple and pure
shear) are summarized in Table 1.

The stress anisotropy measured here is smaller than the value obtained
in other recent work (${\hat \Sigma}_{xy} \approx
0.095$)~\cite{zheng}.  The shear-jamming protocol in this prior work
is very different than the one presented here.  We isotropically
compress the system to a packing fraction below jamming onset for each
particular initial condition, and then apply quasistatic shear at
fixed area until the system first jams at strain $\gamma_J$. In
contrast, in these prior studies, the authors start with jammed
packings at a given pressure $P>0$ and then apply quasistatic shear at
fixed $P$ to a total strain $\gamma=10$.  Thus, the system can undergo
rearrangements and switch from one geometrical family to
another. Moreover, these prior studies only quoted a stress anisotropy
for a finite-sized system ($N=1024$), and did not provide an estimate
for the stress anisotropy in the large system limit.

We will now describe a framework for determining the distribution of
shear stress $P({\hat \Sigma}_{xy})$ for MS packings generated via
protocol 2 with simple shear from the shear stress distribution
obtained from protocol 1. We first make an approximation in
Eq.~\ref{simple_shear}, $\hat{\Sigma}_{xy} \approx
-\frac{1}{\langle \phi \rangle_2}\frac{d \phi}{d \gamma}$, where $\langle \phi \rangle_2$ is the
average packing fraction for MS packings generated using protocol
2. Now, the goal is to calculate the distribution of the dilatancy,
which hereafter we define as $\dot{\phi} \equiv -\frac{d \phi}{d
  \gamma}$.

We first consider an infinitesimal segment of a geometrical family
(labeled $i$) that starts at $(\gamma_i, \phi_i)$ and ends at
$(\gamma_i+\mathrm{d}\gamma, \phi_i-\mathrm{d}\phi)$.  We only need to
consider segments with negative slope, which implies that
$\mathrm{d}\gamma >0$, $\mathrm{d}\phi >0$, and $\dot{\phi}>0$. The
probability to obtain an MS packing on segment $i$ is proportional to
(1) the volume of the initial conditions in configuration space that find
segment $i$~\cite{frenkel,AshwinPRE2012}, $V_{1,i}$ for protocol 1 and
$V_{2,i}$ for protocol 2, and (2) the region of parameter space over
which the segment is sampled, $d\gamma_i$ for protocol 1 and $d\phi_i$
for protocol 2. Thus, $P_{1,i} \propto V_{1,i} d \gamma_i$ for protocol
1 and $P_{2,i} \propto V_{2,i} d \phi_i$ for protocol 2.

The probability distribution for the dilatancy $\dot{\phi}$ can be
written as:
\begin{equation}
\label{eq_pi}
P_{1,2}(\dot{\phi}) = \frac{V_{1,2}(\dot{\phi})}{\int_{0}^{\infty}V_{1,2}(\dot{\phi})\mathrm{d}\dot{\phi}},
\end{equation}
where $V_{1,2}(\dot{\phi})$ is the sum of the basin volumes over all of the infinitesimal segments with slope $\dot{\phi}$,
\begin{subequations}\label{eq_v_seg}
\begin{align}
V_1(\dot{\phi})& = \sum_{i}V_{1,i}(\dot{\phi})\mathrm{d}\gamma_i \label{second}\\
V_2(\dot{\phi})&=  \sum_{i}V_{2,i}(\dot{\phi})\mathrm{d}\phi_i.  \label{third}
\end{align}
\end{subequations}
In the small-$\gamma$ limit
($\gamma_i\approx0$), the basin volumes for each segment $i$
from protocols 1 and 2 satisfy $V_{1,i}
\approx V_{2,i}$. (In Appendix B, we identify the shear strain at which this
approximation breaks down.) In this limit, the protocol dependence
of $P(\dot{\phi})$ is caused by the region of parameter space over which
the MS packings are sampled, $d \gamma_i$ for protocol 1 versus $d \phi_i$
for protocol 2. Thus, the
distribution of dilatancy for protocol 2 for simple shear is given by:
\begin{subequations}
\label{eq_trans}
\begin{align}
P_{2}(\dot{\phi}) &= \frac{\sum_{i}V_{2,i}\mathrm{d}\phi_i}{\int_{0}^{\infty}\sum_{i}V_{2,i}\mathrm{d}\phi_i\mathrm{d}\dot{\phi}} \approx \frac{\sum_{i}V_{1,i}\mathrm{d}\gamma_i\dot{\phi}}{\int_{0}^{\infty}\sum_{i}V_{1,i}\mathrm{d}\gamma_i \dot{\phi}\mathrm{d}\dot{\phi}} \label{eq_trans_a} \\
&\approx \frac{P_{1}(\dot{\phi})\dot{\phi}}{\langle \dot{\phi}\rangle_1}, \label{eq_trans}
\end{align}
\end{subequations}
where we have used the relation $d\phi_i = d\gamma_i {\dot \phi}$ and
$\langle \dot{\phi} \rangle_1$ is the average of $\dot{\phi}$
for MS packings generated using protocol 1 with $\dot{\phi}>0$.

\begin{figure}[h]
\includegraphics[width=8.3 cm]{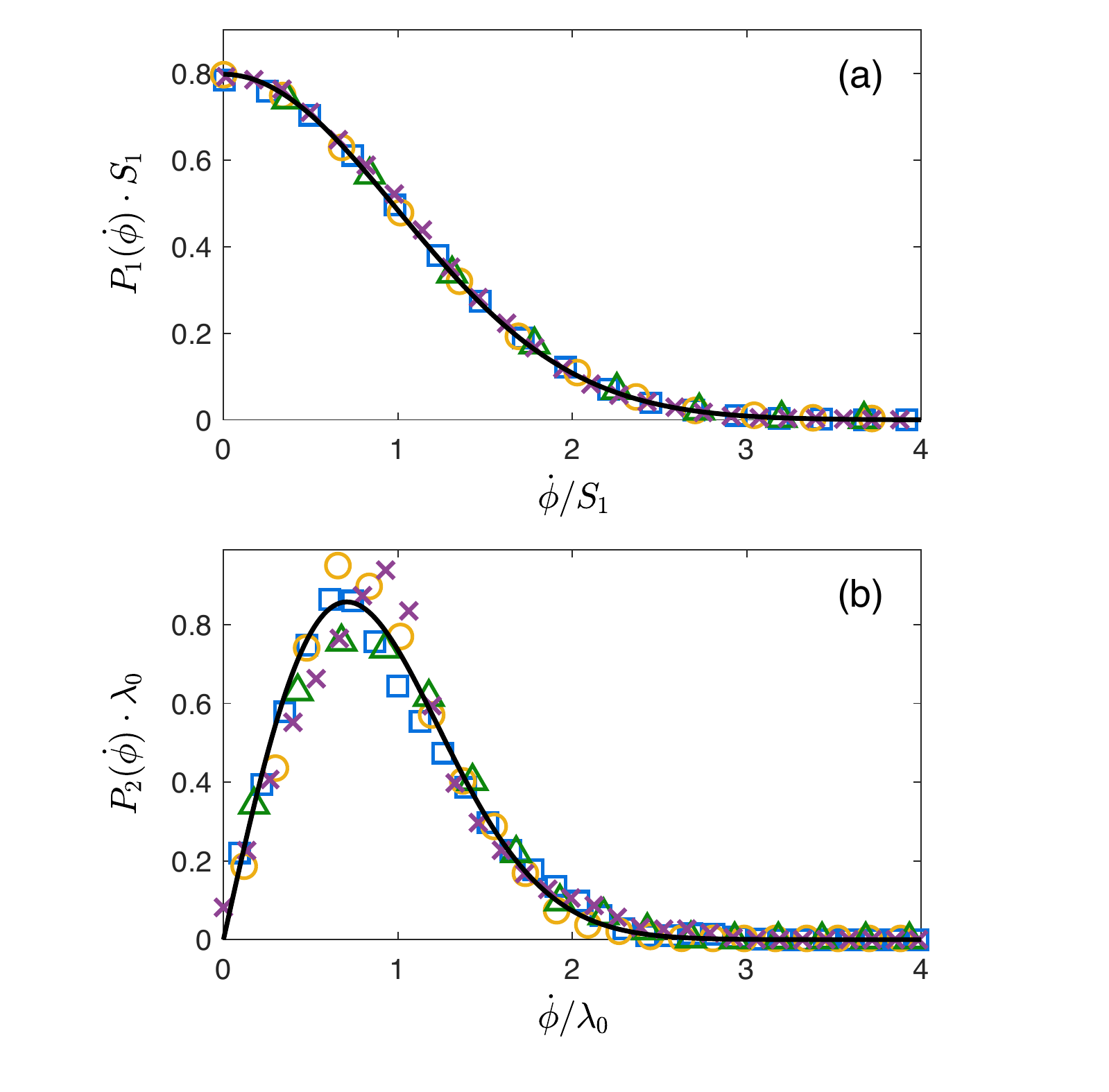}
\caption{\label{fig5_trans} (a) Probability distribution of the
dilatancy $P_1(\dot{\phi})$ for $\dot{\phi} >0$ scaled by the
standard deviation $S_1$ for MS packings generated via protocol 1
with $N = 64$ (squares), $128$ (circles),
$256$ (triangles), and $512$ (crosses). The solid line is the half-Gaussian
distribution in Eq.~\ref{half-gaussian}. (b) Probability distribution
of the dilatancy $P(\dot{\phi})$ for MS packings generated via protocol
2 with simple shear in the small strain limit
($\gamma<2\times10^{-4}$). The symbols are the same as in panel (a). The
solid line is the Weibull distribution in Eq.~\ref{eq_p2gam0} with
shape parameter $k_0 =2$ and scale parameter $\lambda_0 = \sqrt{2}S_1$.}
\end{figure}

In Fig.~\ref{fig4_pdf} (a), we show that the dilatancy distribution
$P_1(\dot{\phi})$ for ${\dot \phi}>0$ from protocol 1 obeys a
half-Gaussian distribution,
%equation P_I (9)
\begin{equation}
\label{half-gaussian}
P_1(\dot{\phi}) = \frac{\sqrt{2}}{S_1\sqrt{\pi} }\mathrm{exp}\left( -\frac{\dot{\phi}^2}{2S_1^2}\right),
\end{equation}
with standard deviation $S_1$. After we substitute $P_{1}(\dot{\phi})$ given by Eq.~\ref{half-gaussian} and $\langle \dot{\phi} \rangle_1 = \sqrt{2/\pi}S_1$
into Eq.~\ref{eq_trans}, we find the following expression for the dilatancy
distribution for MS packings generated via protocol 2 with simple shear in the
small-$\gamma$ limit:
\begin{equation}
\label{eq_p2gam0}
P_{2}(\dot{\phi}|\gamma \ll 1) = \frac{k_0}{\lambda_0}\left(\frac{\dot{\phi}}{\lambda_0}\right)^{k_0-1}\mathrm{exp}\left[ -\left(\frac{\dot{\phi}}{\lambda_0}\right)^{k_0}\right].
\end{equation}
$P_2({\dot \phi}|\gamma \ll 1) = f_w(\dot{\phi};\lambda_0,k_0)$ is a
Weibull distribution with shape parameter $k_0 = 2$ and scale
parameter $\lambda_0 = \sqrt{2}S_1$. We show in Fig.~\ref{fig5_trans}
(b) that the prediction in Eq.~\ref{eq_p2gam0} agrees quantitatively
with the simulation results for $\gamma<2\times10^{-4}$ over a
range of system sizes.

% figure 5

\begin{figure}[h]
\includegraphics[width=8.8 cm]{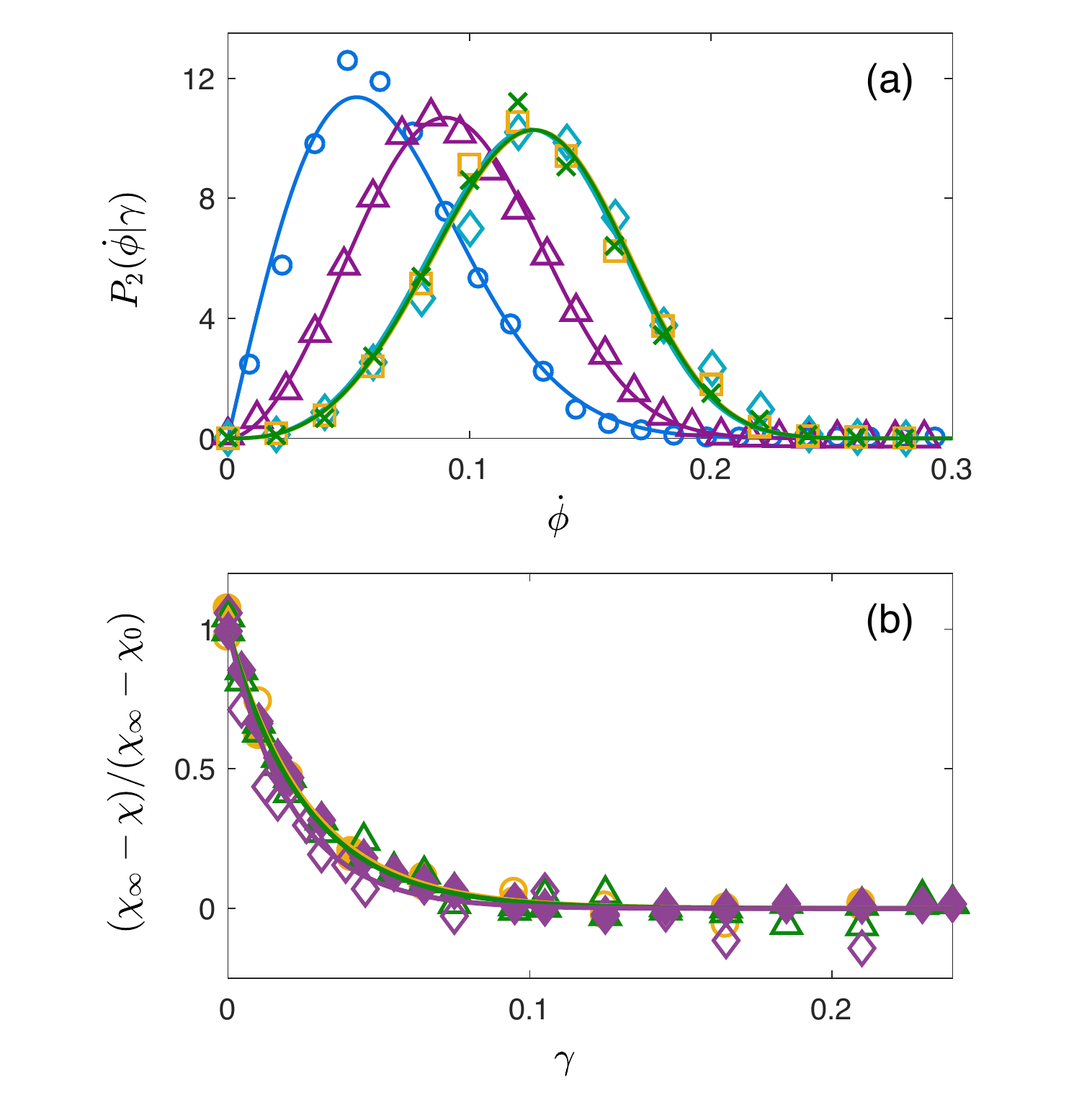}
\caption{
(a) The conditional probability $P_{2}(\dot{\phi}|\gamma)$ for obtaining
dilatancy ${\dot \phi}$ for MS packings with $N=128$ generated
via protocol 2 with simple shear for $\gamma<2\times10^{-4}$
(circles), $0.012<\gamma<0.016$ (triangles), $0.20<\gamma<0.22$
(diamonds), $0.22<\gamma<0.24$ (squares), and $0.24<\gamma<0.26$
(crosses). The solid lines are Weibull distributions $f_w({\dot \phi},\lambda(\gamma),
k(\gamma))$. (b) The $\gamma$-dependence of the shape parameter $\chi=k$ (open symbols)
and scale parameter $\chi=\lambda$ (solid symbols) for fits of
$P_2({\dot \phi}|\gamma)$ to Weibull distributions for $N = 128$ (circles),
$256$ (triangles), and $512$ (diamonds). $\chi_0$ and $\chi_{\infty}$ give
the values of $k$ and $\lambda$ at $\gamma=0$ and in the $\gamma \rightarrow
\infty$ limit, respectively. The solid lines are fits to an exponential
decay, $\sim \exp(-\gamma/\gamma_c)$, where $\gamma_c = 0.027$, $0.026$, and
$0.021$ for $N=128$, $256$, and
$512$, respectively.}
\label{fig_trans2}
\end{figure}

We will now consider the dilatancy distribution for MS packings
generated via protocol 2 at finite shear strains.  For protocol 1
(isotropic compression), our previous studies have shown that the
distribution of jammed packing fractions is independent of the shear
strain $\gamma$~\cite{BertrandPRE2016}. However, for protocol 2
(e.g. with simple shear), systems will preferentially jam on
geometrical families at small $\gamma$, effectively blocking families
at larger $\gamma$, which causes the fraction of unjammed packings to
decay exponentially with increasing $\gamma$ for protocol 2 at a given
$\phi$~\cite{BertrandPRE2016}. Therefore, as $\gamma$ increases, the
assumption that $V_{1,i} \approx V_{2,i}$ is no longer valid, as shown
in Appendix B. To characterize the $\gamma$-dependence of the
dilatancy distribution, we partition the packings into regions of strain
$\gamma$ required to jam them. We can then express the dilatancy
distribution for MS packings generated via protocol 2 with simple
shear as an integral over $\gamma$:
\begin{equation}
\label{eq_gam_int}
P_{2}(\dot{\phi}) = \int_0^{\infty}P_{2}(\dot{\phi}|\gamma)P_{2}(\gamma) d\gamma,
\end{equation}
where $P_{2}(\dot{\phi}|\gamma)$ is the conditional probability for
obtaining $\dot{\phi}$ at a given $\gamma$ and $P_{2}(\gamma)$ is the
probability for obtaining an MS packing as a function of $\gamma$,
which displays exponential decay~\cite{BertrandPRE2016}:
$P_{2}(\gamma)=\alpha \exp(-\alpha\gamma)$.  We show in
Fig.~\ref{fig_trans2} (a) that $P_{2}(\dot{\phi}|\gamma)$ obeys a Weibull
distribution, $f_w(\dot{\phi};\lambda, k)$, with shape
$k(\gamma)$ and scale parameters $\lambda(\gamma)$ that depend on
strain $\gamma$. $k(\gamma)$ and $\lambda(\gamma)$ decay exponentially
to steady-state values in the large-$\gamma$ limit as shown in
Fig.~\ref{fig_trans2} (b):
\begin{equation}
\label{eq_gam_dependence}
\frac{\chi_{\infty}-\chi(\gamma)}{\chi_{\infty}-\chi_0} =  \exp(-\gamma/\gamma_c),
\end{equation}
where $\chi = k$, $\lambda$ and $\chi_0$ and $\chi_{\infty}$ are the
values when $\gamma=0$ and $\gamma \rightarrow \infty$, respectively. We find that both
$k$ and $\lambda$ reach steady-state values when $\gamma > \gamma_c$,
where $\gamma_c \approx 0.02$ in the large-system limit.

%%%%

In the final step, we combine Eqs.~\ref{eq_p2gam0}
and~\ref{eq_gam_int} with the results from Eq.~\ref{eq_gam_dependence}
to predict the distribution of shear stress for MS packings generated
via protocol 2 with simple shear:
\begin{equation}
\label{eq_pdf_prediciton}
P_{2}(\hat{\Sigma}_{xy}) = \langle \phi \rangle_2 \int_0^{\infty} f_w({\dot \phi};\lambda(\gamma),k(\gamma))\alpha \exp \left(-\alpha\gamma\right) d \gamma,
\end{equation}
where $\hat{\Sigma}_{xy} = \dot{\phi}/\langle \phi \rangle_2$ has been used to relate
$P_{2}(\hat{\Sigma}_{xy})$ to $P_{2}(\dot{\phi})$. The results from
Eq.~\ref{eq_pdf_prediciton} agree quantitatively with the distribution
directly calculated from the stress tensor components over a range of
system sizes as shown in Fig.~\ref{fig4_pdf} (a).  Thus, these results
emphasize that we are able to calculate the distribution of shear
stress for MS packings generated via protocol 2
from the distribution of shear stress from MS packings generated via
protocol 1, plus only three parameters: $\alpha \gamma_c$, $k_{\infty}$,
and $\lambda_{\infty}$.  We will show below that $\langle
{\hat \Sigma}_{xy} \rangle$ depends very weakly on $k_{\infty}$.

\subsection{System-size dependence of the average stress anisotropy for
shear-jammed packings}
\label{system_size}

In Fig.~\ref{fig4_pdf}, we showed that the average shear stress
$\langle {\hat \Sigma}_{xy} \rangle \sim 0.06$ reaches a nonzero value
in the large-system limit for MS packings generated via protocol 2
with simple shear.  In this section, we investigate the system
size dependence of $\langle {\hat \Sigma}_{xy} \rangle$ using the
framework (Eq.~\ref{eq_pdf_prediciton}) for calculating the shear
stress distribution for MS packings generated via protocol 2
using the shear stress distribution for MS packings
generated via isotropic compression (protocol 1).

$\langle \hat{\Sigma}_{xy} \rangle$ for MS packings generated
via protocol 2 can be calculated from the probability distribution $P_2({\hat
\Sigma}_{xy})$:
\begin{equation}
    \label{eq_mean_pre1}
  \begin{aligned}
\langle \hat{\Sigma}_{xy}\rangle &= \int_{0}^{\infty} \hat{\Sigma}_{xy} P_2(\hat{\Sigma}_{xy}) d\hat{\Sigma}_{xy} \\
                      &\approx \int_{0}^{\infty}  \frac{\dot{\phi}}{\langle \phi \rangle_2} (\langle \phi \rangle_2 P_2(\dot{\phi}))\frac{\mathrm{d}\dot{\phi}}{\langle \phi \rangle_2} = \frac{1}{\langle \phi \rangle_2}  \int_0^{\infty}\dot{\phi}P_2(\dot{\phi}) \mathrm{d}\dot{\phi}.
\end{aligned}
\end{equation}

\begin{figure}
\includegraphics[width = 8.5 cm]{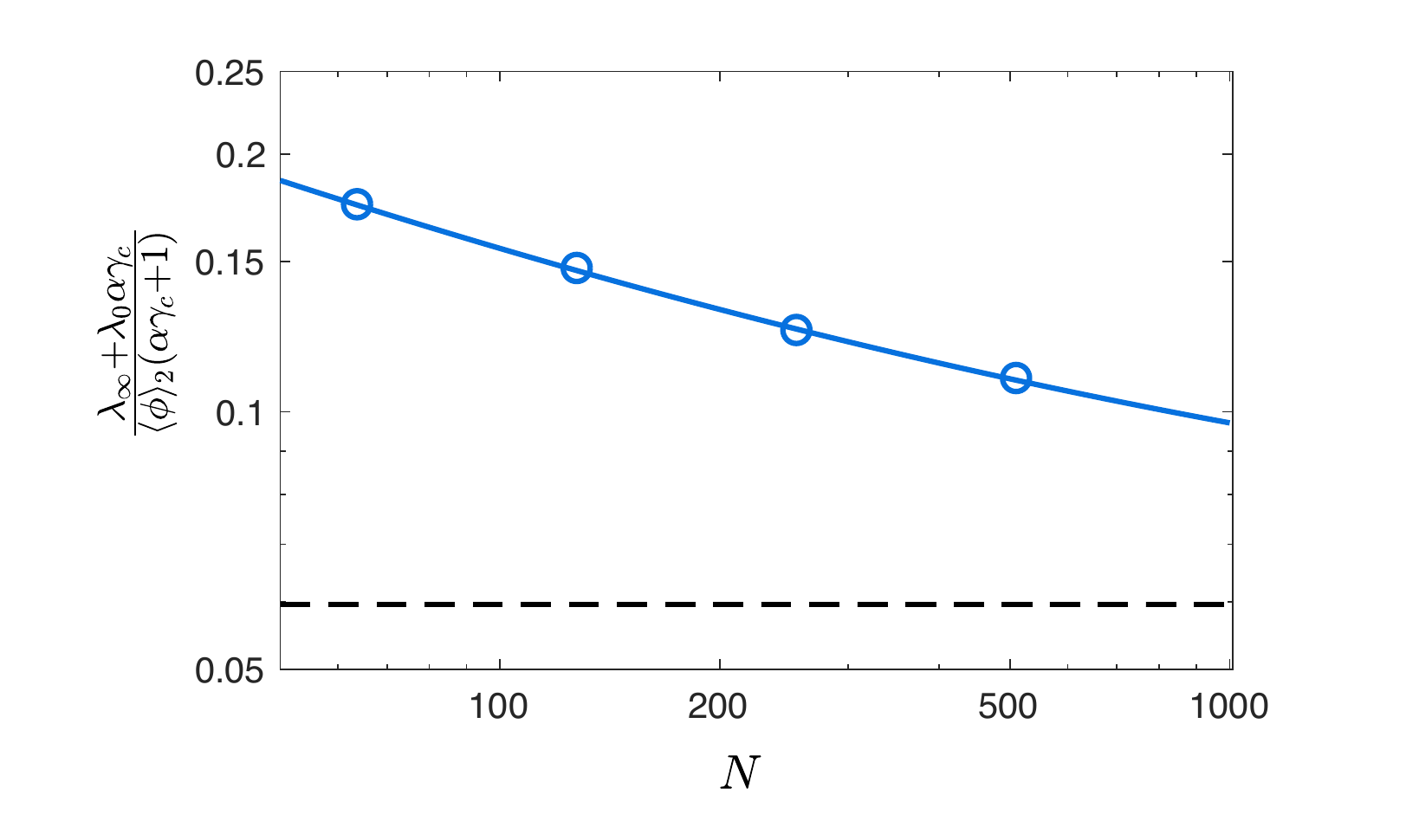}
\caption{The system-size dependence of $\langle {\hat \Sigma}_{xy} \rangle
\approx (\lambda_{\infty} + \lambda_0 \alpha \gamma_c )/[\langle \phi \rangle_2(\alpha \gamma_c+1)]$ (circles) from
Eq.~\ref{eq_sigma_app2}. The best fit to Eq.~\ref{eq_fit_nddep} is
given by the solid line. The shear stress in the large-system limit $\langle {\hat \Sigma}_{xy}
\rangle_{\infty} \approx 0.060$ is indicated by the dashed line.}
  \label{fig_Ndep}
\end{figure}

After substituting Eq.~\ref{eq_gam_int} into Eq.~\ref{eq_mean_pre1}, we have
\begin{equation}
  \label{eq_mean_pre2}
  \begin{aligned}
  \langle \hat{\Sigma}_{xy}\rangle &=  \frac{1}{\langle \phi \rangle_2}  \int_0^{\infty}\dot{\phi} \left( \int_0^{\infty} f_w(\dot{\phi};\lambda(\gamma),k(\gamma))\alpha
\exp \left(-\alpha\gamma\right) d \gamma \right)
  d \dot{\phi}\\
  &=\frac{1}{\langle \phi \rangle_2} \int_0^{\infty} \langle \dot{\phi}\rangle_{\gamma}\alpha
\exp \left(-\alpha\gamma\right) d \gamma,
\end{aligned}
\end{equation}
where $\langle \dot{\phi} \rangle_{\gamma} =
\lambda(\gamma)\Gamma(1+1/k(\gamma))$ is the average of
$\dot{\phi}$ at strain $\gamma$. The shape parameter $k(0)=2$ and
increases with $\gamma$, and thus $0.886 \lesssim \Gamma(1 + 1/k(\gamma)) <
1$. Therefore, $\langle \dot{\phi}
\rangle_{\gamma}$ can be approximated as
\begin{equation}
  \label{eq_sigma_app1}
\langle \dot{\phi}\rangle_{\gamma} \approx \lambda(\gamma) = \lambda_{\infty}[1-\exp(-\gamma/\gamma_c)]+\lambda_0 \exp(-\gamma/\gamma_c).
\end{equation}
After substituting Eq.~\ref{eq_sigma_app1} into Eq.~\ref{eq_mean_pre2}, we
find
\begin{equation}
\label{eq_sigma_app2}
\langle \hat{\Sigma}_{xy}\rangle \approx \frac{\lambda_{\infty} + \lambda_0 \alpha \gamma_c }{\langle \phi \rangle_2(\alpha \gamma_c + 1)},
\end{equation}
which is plotted versus system size in
Fig.~\ref{fig_Ndep}. We fit the system-size dependence to following
form:
\begin{equation}
\label{eq_fit_nddep}
\langle {\hat \Sigma}_{xy} \rangle(N) = {\hat \Sigma}_0 N^{-\Omega} + {\hat
\Sigma}_{\infty},
\end{equation}
where ${\hat \Sigma}_0 \approx 0.62$, $\Omega \approx 0.41$, and
${\hat \Sigma}_{\infty} \approx 0.060$, which are similar to the
values found directly using the data in Fig.~\ref{fig4_pdf}.

\section{Conclusions and Future Directions}
\label{conclusions}

In this article, we carried out computer simulations of frictionless,
purely repulsive disks to investigate the
development of stress anisotropy in mechanically stable (MS) packings
prepared using two protocols. Protocol 1 involves shearing the
system quasistatically to a given strain at low packing
fraction and then compressing the system quasistatically to jamming
onset at fixed strain. Protocol 2 involves compressing the
system quasistatically at $\gamma=0$ to a packing fraction
below jamming onset, and then shearing the system quasistatically to
achieve jamming onset.

We find that the stress anisotropy distribution for MS packings generated via protocol 1 is a Gaussian with zero mean and a standard deviation that scales to zero in the large-system limit.  In contrast, MS packings prepared using protocol 2 have a nonzero stress anisotropy ${\hat \tau}_{\infty} \approx 0.06$ and standard deviation $S_{\infty} \approx 0.015$ in the large-system limit. We also derived relationships between the components of the stress tensor (shear stress and normal stress difference) and the dilatancy $d\phi/d\gamma$. Using these relations, we developed a statistical
framework to calculate the stress anisotropy distribution for
shear-jammed packings (i.e. MS packings generated via protocol 2) in
terms of the stress anisotropy distribution for isotropically prepared
packings (i.e. MS packings generated via protocol 1). We showed that
the stress anisotropy distribution for shear-jammed packings can be
described by a convolution of Weibull distributions with shape and
scale parameters that depend on strain. The results for the stress
anisotropy distribution from the statistical framework agree
quantitatively with the direct measurements of the stress tensor for
MS packings generated using protocol 2. These results emphasize that
the packing-generation protocol can dramatically influence the
probabilities with which MS packings occur, and thus change the
average macroscopic quantities that are measured for a given protocol.

There are several interesting directions for future research
investigating the development of stress anisotropy in jammed
systems. First, how does the presence of frictional interparticle
forces affect this picture?  Recent computational studies have shown
that the shear modulus displays a discontinuous jump with increasing
strain for static packings of frictional spheres~\cite{otsuki}.  Can
the discontinuity in the shear modulus be explained using the
statistical framework for the shear stress distribution that we
developed here?  Moreover, there are still open questions about
whether pure/simple shear and isotropic compression can give rise to
fundamentally different ensembles of MS packings of frictional
particles. For example, consider the Cundall-Strack model for static
friction between contacting grains~\cite{cundall}. In this model, the tangential
force, which is proportional to the relative tangential displacement
between contacting grains can grow until the ratio of the magnitude of
the tangential to normal force reaches the static friction coefficient
$\mu$.  If the ratio exceeds $\mu$, the particle slips and the
relative tangential displacement is reset.  Two packings with
identical particle positions can possess different numbers of
near-slipping contacts. It is thus possible that different
packing-generation protocols will lead to nearly identical MS packings
with different numbers of near-slipping contacts.

Second, how does non-spherical particle shape affect the geometrical
families $\phi(\gamma)$?  In preliminary studies, we have shown that
the geometrical families for MS packings of circulo-polygons occur as
parabolic segments that are both concave up and concave down.  (See
Appendix C.)  In future studies, we will generate packings of
circulo-polygons using protocol 2 to connect the statistics of the
geometrical families $\phi(\gamma)$ to the development of nonzero
stress anisotropy in the large-system limit for MS packings of
non-spherical particles.

\section*{Appendix A: Normal stress difference ${\hat \Sigma}_N$ for
MS packings generated
via protocol 2 with pure shear}

In Fig.~\ref{fig4_pdf}, we presented the probability distributions for
the shear stress $\hat{\Sigma}_{xy}$ and normal stress difference
$\hat{\Sigma}_{N}$ for MS disk packings generated via protocol 1 and
protocol 2 with simple shear. In this Appendix, we show the
results for the probability distributions $P({\hat \Sigma}_{xy})$ and
$P({\hat \Sigma}_N)$ for MS disk packings generated via
protocol 2 with pure shear.

\begin{figure}[h]
\includegraphics[width=8.0 cm]{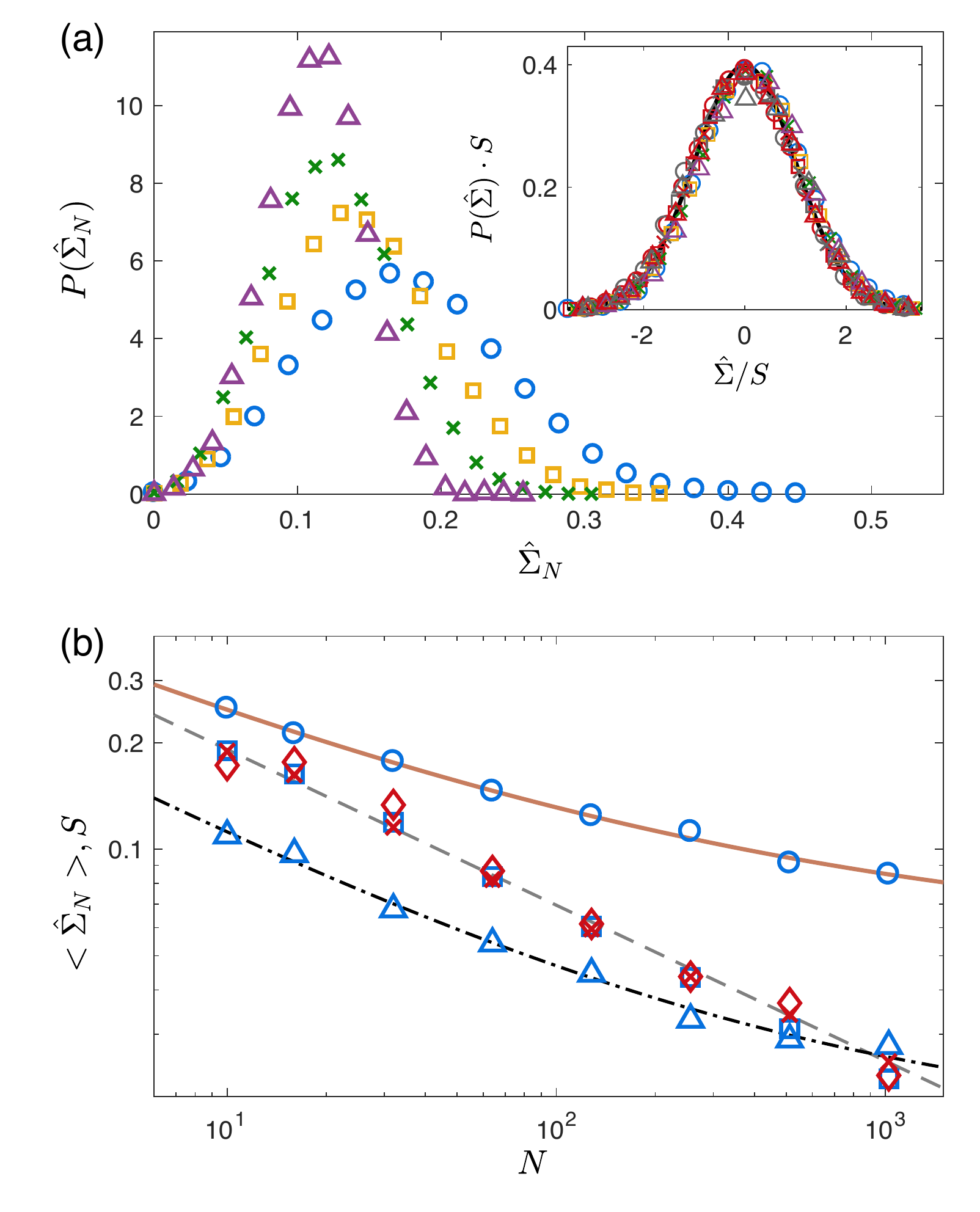}
\caption{\label{fig_pdf_pure}
(a) The probability distribution $P(\hat{\Sigma}_{N})$ of the normal
stress difference for MS packings generated via protocol 2 with pure shear
for $N = 32$ (circles), $64$ (squares), $128$ (crosses), and $256$
(triangles). The inset shows the distributions
for ${\hat \Sigma}_{N}$ (same symbols as in main panel) and
$\hat{\Sigma}_{xy}$ (same symbols as in main panel, but
in red) for MS
packings generated via protocol 1, and $\hat{\Sigma}_{xy}$ for MS packings generated via
protocol 2 using pure shear
(same symbols as in main panel, but in gray). The solid black line is a
Gaussian distribution with a zero mean and unit standard deviation. (b)
System-size dependence of 1) the average (circles) and standard deviation
(triangles) of $P({\hat \Sigma}_{N})$ for
MS packings generated via protocol 2 with pure shear, 2)
standard deviation of $P({\hat \Sigma}_{xy})$ (squares) for
MS packings generated via protocol 2 with pure shear, and 3)
standard deviations of $P({\hat \Sigma}_{N})$
(crosses) and $P({\hat \Sigma}_{xy})$ (diamonds) for MS packings
generated via protocol 1.
The dashed, solid, and dash-dotted lines are fits to
Eqs.~\ref{s_cale_p1},~\ref{pure_normmean}, and~\ref{pure_normstd},
respectively.}
\end{figure}

Pure shear strain couples to the normal stress difference, not to the
shear stress. Thus, as shown in Fig.~\ref{fig_pdf_pure} (a), the
probability distributions $P({\hat \Sigma}_N)$ for MS packings
generated via protocol 2 with pure shear are qualitatively the same as
$P({\hat \Sigma}_{xy})$ for MS packings generated via protocol 2 with
simple shear. The probability distributions $P({\hat \Sigma}_{N})$ and
$P({\hat \Sigma}_{xy})$ for MS packings generated via protocol 1
and $P({\hat \Sigma}_{xy})$ for MS packings generated via protocol 2 (with
pure shear)
are Gaussian with zero mean and standard deviations that scale to
zero with increasing system size. (See Eq.~\ref{s_cale_p1}.)

%fig pdf pure shear figure 9

The average of $P({\hat \Sigma}_{N})$ for MS packings generated
via protocol 2 with pure shear decreases as $N$ increases, but reaches
a nonzero value in the large-system limit:
\begin{equation}
  \label{pure_normmean}
  \langle \hat{\Sigma}_{N}\rangle (N) = \hat{\Sigma}_0 N^{-\Omega}+\hat{\Sigma}_{\infty},
\end{equation}
where $\hat{\Sigma}_0 \approx 0.49$, $\Omega \approx 0.40$, and
$\hat{\Sigma}_{\infty} \approx 0.055$. Similarly, the standard deviation of
$P({\hat \Sigma}_{N})$ also reaches a nonzero value in the large-system
limit:
\begin{equation}
  \label{pure_normstd}
  S_2(N) = S_2^0 N^{-\omega_2}+S_{\infty},
\end{equation}
where $S_2^0 \approx 0.30$, $\omega_2 \approx 0.50$, and
$S_{\infty} \approx 0.016$.  The results for MS packings generated via
protocol 2 with pure shear are analogous to those observed for MS packings
generated via protocol 2 with simple shear. (See Table~1.)

\begin{figure*}
\includegraphics[width=14 cm]{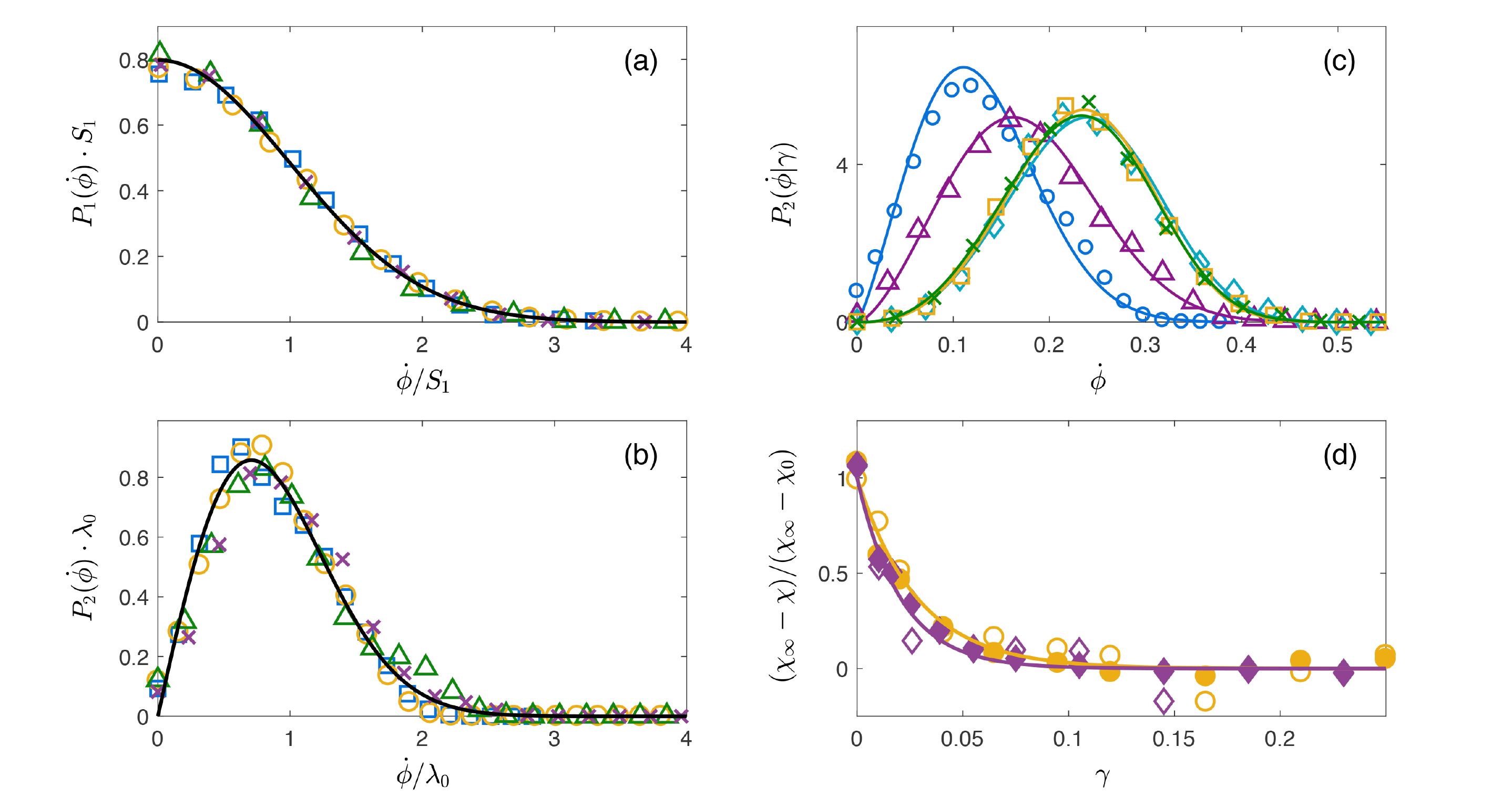}
\centering
\caption{\label{fig_pure}
(a) Probability distribution of the
dilatancy $P_1(\dot{\phi})$ for $\dot{\phi} >0$ scaled by the
standard deviation $S_1$ for MS packings generated via protocol 1 and pure shear
with $N = 64$ (squares), $128$ (circles),
$256$ (triangles), and $512$ (crosses). The solid line is the half-Gaussian
distribution in Eq.~\ref{half-gaussian}. (b) Probability distribution
of the dilatancy $P(\dot{\phi})$ for MS packings generated via protocol
2 with pure shear in the small strain limit
($\gamma<2\times10^{-4}$). The symbols are the same as in panel (a). The
solid line is the Weibull distribution in Eq.~\ref{eq_p2gam0} with
shape parameter $k_0 =2$ and scale parameter $\lambda_0 = \sqrt{2}S_1$.
(c) The conditional probability $P_{2}(\dot{\phi}|\gamma)$ for obtaining
dilatancy ${\dot \phi}$ for MS packings with $N=128$ generated
via protocol 2 with pure shear for $\gamma<2\times10^{-4}$
(circles), $0.010<\gamma<0.0105$ (triangles), $0.18<\gamma<0.20$
(diamonds), $0.20<\gamma<0.22$ (squares), and $0.24<\gamma<0.26$
(crosses). The solid lines are Weibull distributions $f_w({\dot \phi},\lambda(\gamma),
k(\gamma))$. （d) The $\gamma$-dependence of the shape parameter $\chi=k$ (open symbols)
and scale parameter $\chi=\lambda$ (solid symbols) for fits of
$P_2({\dot \phi}|\gamma)$ to Weibull distributions for $N = 128$ (circles) and $512$ (diamonds). $\chi_0$ and $\chi_{\infty}$ give
the values of $k$ and $\lambda$ at $\gamma=0$ and in the $\gamma \rightarrow
\infty$ limit, respectively. The solid lines are fits to an exponential
decay, $\sim \exp(-\gamma/\gamma_c)$, where $\gamma_c = 0.029$ and
$0.021$ for $N=128$ and
$512$, respectively.}
\end{figure*}

MS packings generated via protocol 2 for pure shear obey the same
stress-dilatancy relationship (Eq.~\ref{simple_shear}) as that for
simple shear. Thus, we can apply the statistical model in
Sec. ~\ref{stress_anisotropy} to predict the stress anisotropy
distribution for MS packings generated via pure shear. As shown in
Fig. ~\ref{fig_pure} (a) and (b), the distribution for the dilatancy
of shear-jammed packings at small $\gamma$ limit obeys a Weibull
distribution, which can be predicted from the half-Gaussian
distribution for MS packings obtained via protocol 1. (See
Eqs. ~\ref{half-gaussian} and ~\ref{eq_p2gam0}.)  The conditional
probability, $P_2(\dot{\phi}|\gamma)$, for obtaining $\dot{\phi}$ at
a given $\gamma$ is shown in Fig. ~\ref{fig_pure} (c) and fit to a
Weibull distribution $f_w(\dot{\phi};\gamma,k)$. In
Fig. ~\ref{fig_pure} (d), we plot the $\gamma$ dependence of the
shape $k(\gamma)$ and scale $\lambda(\gamma)$ parameters. Both
parameters decay exponentially to steady-state values in the
large-$\gamma$ limit. (See Eq.~\ref{eq_gam_dependence}.) These
results are similar to those for simple shear case described in the
main text.

\section*{Appendix B: Protocol dependence of the volume of the basin of
attraction for MS packings}
\label{appc}

In the description of the statistical framework
(Sec.~\ref{stress_anisotropy}) for calculating the distribution of
dilatancy for MS packings generated via protocol 2 with simple shear
from those generated via protocol 1, we first assumed that the volumes
of the basins of attraction were the same (i.e. $V_{1,i}\approx
V_{2,i}$) for protocols 1 and 2.  In this Appendix, we illustrate
that this assumption breaks down for sufficiently large simple shear strains.

\begin{figure}[h]
\includegraphics[width=8.5 cm]{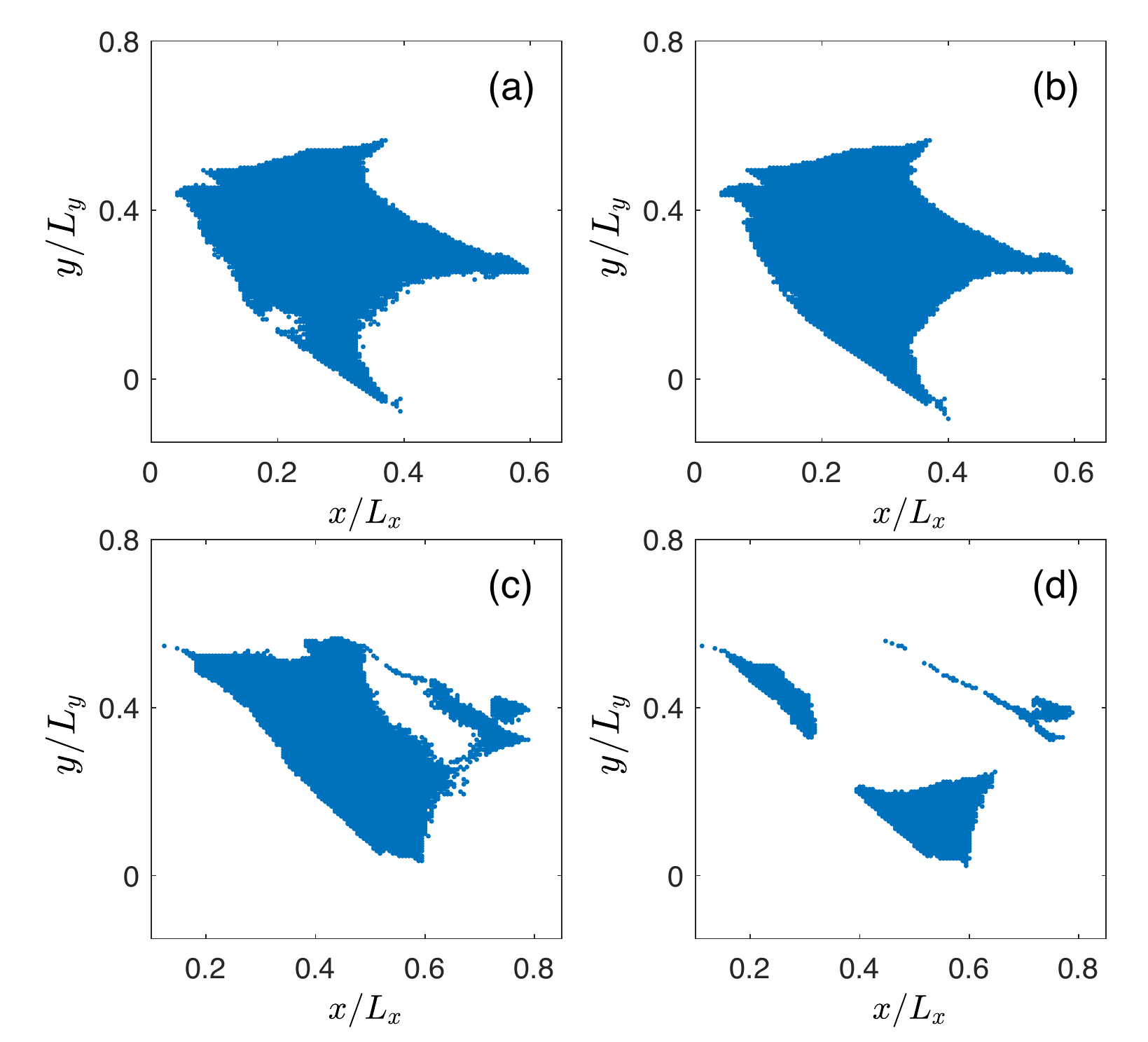}
\caption{\label{fig_s_basin}
Two-dimensional projection of the four-dimensional basin of attraction
for a particular $N=6$ MS packing generated using (a) protocol 1 and
(b) protocol 2 (with simple shear) at shear strain $\gamma = 2\times 10^{-3}$.  The
$x$- and $y$-coordinates indicate the initial position of particle $1$,
while the initial conditions for particles $2$ through $6$ are fixed to specific locations within
the simulation cell. If a pixel is blue, the initial position maps
to the target MS packing after the packing-generation procedure. Panels (c) and (d) are comparable to (a) and (b)
except the shear strain has been increased to $\gamma=0.02$.}
\end{figure}

We illustrate the basin volume for an $N=6$ MS packing, which is a
four-dimensional quantity, by projecting it into two dimensions.  We
consider a particular $N=6$ MS packing at shear strain $\gamma$ and
packing fraction $\phi$ that can be generated readily via protocol 1 and
protocol 2 with simple shear.  We identify a point $(\mathbf{r}_1,
\mathbf{r}_2,\ldots,\mathbf{r}_6)$ within the basin of attraction of
the MS packing and constrain the positions of particles $2$ through
$6$. The initial position of particle $1$ is allowed to vary in the
$x$-$y$ plane. The pixels in each panel of Fig.~\ref{fig_s_basin}
represent the initial positions of particle $1$ and they are colored blue if
the initial configuration at ($x$,$y$) maps to the position of particle $1$
in the particular MS packing that we selected. The area of the blue
region gives the projected area of the basin of attraction for that
particular MS packing.

In Fig.~\ref{fig_s_basin} (a) and (b), we show the basins of
attraction for a particular MS packing at a small shear strain,
$\gamma = 2 \times 10^{-3}$, for protocols 1 and 2,
respectively.  The areas of the blue regions are nearly the same,
which suggests that $V_{1,i} \approx V_{2,i}$. However, at larger
shear strains, the basin volumes for the two protocols deviate. For
example, in Fig.~\ref{fig_s_basin} (c) and (d) at shear strain
$\gamma = 0.02$, the projected area for protocol 1 is much larger than that
for protocol 2, which implies
that $V_{1,i} \ne V_{2,i}$.

\section*{Appendix C: Simple shear of circulo-triangle packings}
\label{appd}

In this Appendix, we show that MS packings of non-spherical particles,
specifically circulo-triangles, also form geometrical families in the
packing fraction $\phi$ and shear strain $\gamma$ plane. We considered
bidisperse mixtures of circulo-triangles, half large and half small
with area ratio $r_a = 1.4^2$ and interior angles of $33^{\circ}$,
$62^{\circ}$, and $85^{\circ}$ for each triangle.  We fixed the asphericity parameter
$\mathcal{A} = p^2/4\pi a=1.1$, where $p$ and $a$ are
the perimeter and area of the circulo-triangles, respectively. At this asphericity,
the packings can be either isostatic or hypostatic~\cite{vanderwerf}.

As is the case for circular disks, we find that the geometrical
families for MS packings of circulo-triangles generated via protocol 1
with simple shear form parabolic segments in the $\phi$-$\gamma$
plane, satisfying $\phi(\gamma) = A(\gamma - \gamma_0)^2 +
\phi_0$. However, we find that the curvature of the parabolas can be
both concave up and concave down ($A > 0$ and $A < 0$) for MS packings
of circulo-triangles. In contrast, $A>0$ for MS disk packings.  $A<0$
implies strain-induced compaction, which may be caused by the
alignment of the circulo-triangles during shear.  Preliminary results
indicate that the stress anisotropy for shear jammed packings of
circulo-triangles is finite (and larger than that for frictionless
disks) in the large system limit.

\begin{figure}[h]
\includegraphics[width=7.5 cm]{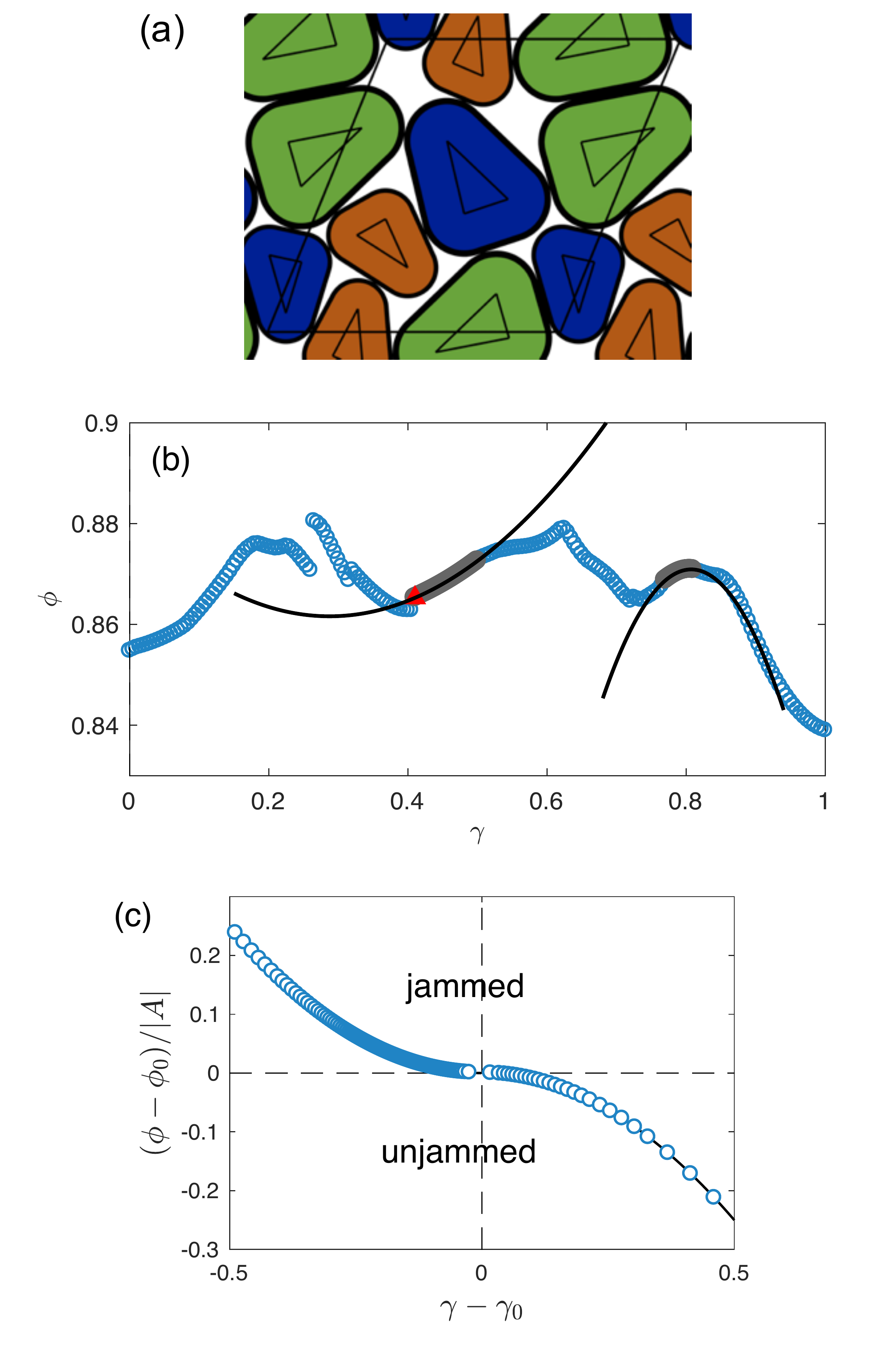}
\caption{\label{fig_s_tri}
(a) An MS packing of $N=6$ of bidisperse circulo-triangles with asphericity
parameter ${\cal A}=1.1$. (b) Packing fraction $\phi$ at jamming onset
as a function of simple shear strain
$\gamma$ for $N=6$ MS packings of circulo-triangles generated via protocol 1.
The packing in panel (a) corresponds to the filled red triangle.
The solid lines are fits of two particular parabolic regions (shaded gray) to
$\phi(\gamma) = A(\gamma - \gamma_0)^2 + \phi_0$. (c) $(\phi-\phi_0)/|A|$
versus $\gamma-\gamma_0$, for $N=6$ MS packings of circulo-triangles
generated via protocol 2 with simple shear. These packings populate the
parabolic regions with $d\phi/d\gamma < 0$ on segments with both $A>0$
and $A<0$. The jammed and unjammed regions
of the $(\phi-\phi_0)/|A|$ and $\gamma-\gamma_0$ plane are indicated.}
\end{figure}

\section*{Acknowledgements}
We acknowledge support from NSF Grants No. CMMI-1462439 (C.O.),
No. CMMI-1463455 (M.S.), and No. CBET-1605178 (C.O.) and China
Scholarship Council No. 201606210355 (S.C.) and No. 201606010264
(W.J.). This work also benefited from the facilities and staff of the
Yale University Faculty of Arts and Sciences High Performance
Computing Center. We thank A. Boromand, A. Clark, K. VanderWerf, and
S. Li for their helpful comments.

CSO, MDS, and TB designed the research, S. Chen and W. Jin performed
the simulations, and S. Chen developed the statistical description.

%%%END OF MAIN TEXT%%%

%The \balance command can be used to balance the columns on the final page if desired. It should be placed anywhere within the first column of the last page.

\balance

%If notes are included in your references you can change the title from 'References' to 'Notes and references' using the following command:
%\renewcommand\refname{Notes and references}

%%%REFERENCES%%%
\bibliography{rsc_sa} %You need to replace "rsc" on this line with the name of your .bib file
\bibliographystyle{rsc} %the RSC's .bst file

\end{document}